\documentclass[a4paper,fleqn,usenatbib]{mnras}

\usepackage{newtxtext,newtxmath}

\usepackage[T1]{fontenc}
\usepackage{ae,aecompl}

\usepackage{graphicx}	
\usepackage{amsmath}	
\usepackage{amssymb}	

\newcommand{\angstrom}{\textup{\AA}}

\title[H$_2$ and CO in galaxy simulation]{Populating H$_2$ and CO in galaxy simulation with dust evolution} 

\author[L.-H. Chen et al.]{ 
Li-Hsin Chen,$^{1,2}$
Hiroyuki Hirashita,$^{1}$
Kuan-Chou Hou,$^{1,3}$
Shohei Aoyama$^{1}$
\newauthor 
Ikkoh Shimizu,$^{4}$
and Kentaro Nagamine$^{4,5}$
\\
$^{1}$Institute of Astronomy and Astrophysics, Academia Sinica, PO Box 23-141, Taipei 10617, Taiwan\\
$^{2}$Department of Physics, National Taiwan University, Taipei 10617, Taiwan\\
$^{3}$Department of Physics $\&$ Institute of Astrophysics, National Taiwan University, Taipei 10617, Taiwan\\
$^{4}$Theoretical Astrophysics, Department of Earth $\&$ Space Science, Osaka University, 1-1 Machikaneyama, Toyonaka, Osaka 560-0043, Japan\\
$^{5}$Department of Physics $\&$ Astronomy, University of Nevada, Las Vegas, 4505 S. Maryland Pkwy, Las Vegas, NV 89154-4002, USA
}

\date{Accepted XXX. Received YYY; in original form ZZZ}

\pubyear{2017} 

\begin{document}
\label{firstpage}
\pagerange{\pageref{firstpage}--\pageref{lastpage}}
\maketitle

\begin{abstract}
There are two major theoretical issues for the star formation law (the relation between the surface densities of molecular gas and star formation rate on a galaxy scale): (i) At low metallicity, it is not obvious that star-forming regions are rich in H$_2$ because the H$_2$ formation rate depends on the dust abundance; and (ii) whether or not CO really traces H$_2$ is uncertain, especially at low metallicity. To clarify these issues, we use a hydrodynamic simulation of an isolated disc galaxy with a spatial resolution of a few tens parsecs. The evolution of dust abundance and grain size distribution is treated consistently with the metal enrichment and the physical state of the interstellar medium. We compute the H$_2$ and CO abundances using a subgrid post-processing model based on the dust abundance and the dissociating radiation field calculated in the simulation. We find that when the metallicity is $\lesssim 0.4$ Z$_\odot$ ($t<1$ Gyr), H$_2$ is not a good tracer of star formation rate because H$_2$-rich regions are limited to dense compact regions. At $Z\gtrsim 0.8$ Z$_\odot$, a tight star formation law is established for both H$_2$ and CO. At old ($t \sim 10$ Gyr) ages, we also find that adopting the so-called MRN grain size distribution with an appropriate dust-to-metal ratio over the entire disc gives reasonable estimates for the H$_2$ and CO abundances. For CO, improving the spatial resolution of the simulation is important while the H$_2$ abundance is not sensitive to sub-resolution structures at $Z\gtrsim 0.4$ Z$_\odot$.
\end{abstract}




\section{Introduction}
Stars form in molecular clouds; thus, understanding how molecular clouds form is an important step to clarify how stars form as the galaxy evolves. H$_2$ is the most abundant component in molecular clouds, but is observationally difficult to detect because of the lack of emission. Detection of CO emission is often used as a tracer of H$_2$ through the CO-to-H$_2$ conversion factor \citep[e.g.][]{Bolatto:2013aa}.

H$_2$ forms more efficiently on dust grain surfaces than in the gas phase \citep[e.g.][]{Gould:1963aa,Cazaux:2004aa}. On the other hand, CO forms through several reaction paths in the gas phase \citep[e.g.][]{Glover:2010aa}. Both molecules are dissociated by ultraviolet (UV) radiation. At high column densities, H$_2$ can shield UV radiation by their own absorption \citep[i.e. self-shielding;][]{Draine:1996aa}. Self-shielding of CO is weaker; therefore, CO formation needs a more embedded environment \citep{Wolfire:2010aa,Shetty:2011aa}. Dust extinction is also important in shielding UV radiation.

Observationally, a clear relation between the surface densities of molecular gas and star formation rate (SFR) for nearby galaxies has been shown to exist \citep{Kennicutt:1998aa, Bigiel:2008aa, Rahman:2012aa, Leroy:2013aa}. In this paper, we refer to the relation between the surface densities of molecular gas mass and SFR as the star formation law. The star formation law is important in understanding how efficiently molecular gas is converted into stars in galaxies  and in constraining theoretical models of star formation \citep[e.g.][]{Semenov:2016aa}. However, it is not clear whether this relation holds at any stage of galaxy evolution. In particular, the CO-to-H$_2$ conversion factor has been shown to depend on metallicity \citep[][for a review]{Bolatto:2013aa}.

The dust abundance is important for both H$_2$ and CO abundances through UV shielding and H$_2$ formation. The grain size distribution also plays an important role in determining the abundances of those molecules since shielding and H$_2$ formation both depend on the total grain surface area \citep[][hereafter HH17]{Yamasawa:2011aa, Hirashita:2017aa}. Therefore, we need to clarify the evolution of both dust abundance and grain size distribution simultaneously to understand the evolution of H$_2$ and CO abundances.

Dust is mainly produced by stellar sources, especially supernovae (SNe) at the early stage of galaxy evolution \citep{Todini:2001aa,Nozawa:2007aa,Bianchi:2007aa} \citep[see Section 2.1 of][for detailed references]{Hirashita:2015aa}. Dust grains produced by stellar sources are considered to be biased to large sizes ($\gtrsim 0.1~\micron$). As the galaxy evolves, the dust abundance in the ISM becomes high enough to activate interstellar processing. Among the interstellar processing mechanisms, shattering and accretion enhance the abundance of small grains, SN destruction decreases the abundances of both large and small grains, and coagulation increases large grains \citep{Asano:2013aa}. Thus, dust abundance is dominated by large grains at the early stage and the abundance of small grains increases as the galaxy evolves.

Numerical simulations have been used to gain insight on dust formation and its impact on galaxies. For example, \citet{Yajima:2014aa} assumed constant dust-to-metal ratio and studied the impact of silicate versus graphite dust on the properties of Lyman break galaxies at $z\sim 3$. \citet{Bekki:2015ab} implemented dust evolution in an $N$-body/smoothed particle hydrodynamics (SPH) galaxy-scale simulation. They considered dust growth by accretion of gas-phase metals in the ISM and destruction by SNe as well as dust production by stellar sources. \citet{Zhukovska:2016aa} developed a 3-dimensional model and used hydrodynamic simulations to investigate dust evolution of giant molecular clouds in a Milky-Way like galaxy. \citet{McKinnon:2016aa} developed a dust model in a hydrodynamic galaxy evolution simulation by considering the same processes as  above and broadly reproduced the relations between dust mass and gas/stellar mass. In addition to that, \citet{McKinnon:2017aa} performed a cosmological simulation and reproduced statistical properties of dust abundance in galaxies at $z \sim 0$. There are also semi-analytic models that incorporated dust evolution in a cosmological context, focusing on the Milky Way progenitors \citep{de-Bennassuti:2014aa}, nearby galaxies \citep{Ginolfi:2017aa} or statistical properties at high redshift \citep{Popping:2017aa}.

There are some simulation efforts incorporating the evolution of grain size distribution. \citet[hereafter A17]{Aoyama:2017aa} treated the evolution of grain size distribution by adopting the 'two-size approximation' by \citet{Hirashita:2015aa} to represent the grain size distribution. \citet{Hirashita:2015aa} divided dust grains into small and large grains around  $a \sim$ 0.03 $\mu$m ($a$ is the grain radius) and showed that this two-size approximation reproduces the same behaviour of a full treatment of grain size distribution by \citet{Asano:2013aa}. In fact, the two-size approximation has been applied to modelling of molecular gas content. HH17 adopted the two-size approximation and examined the effect of dust evolution on H$_2$ and CO formation in clouds with hydrogen column density $N_\mathrm{H} = 10^{21}$--$10^{23}$ cm$^{-2}$. They confirmed the importance of grain size distribution on the H$_2$ and CO abundances and reproduced the observed $X_\mathrm{CO}$--metallicity relation. However, because their model was based on a one-zone treatment, they were not able to predict spatial distribution of molecules. 

In this paper, to further understand the dependence of H$_2$ and CO formation on dust evolution under different physical conditions in a galaxy disc, we adopt a hydrodynamic simulation of an isolated disc galaxy with dust evolution included from A17. We choose a few snapshots at different evolutionary stages of the simulated galaxy and calculate the H$_2$ abundance by assuming the equilibrium between formation and destruction. We also implement a subgrid model to calculate the CO abundance utilizing a method used by \citet{Feldmann:2012aa}. We are particularly interested in how well the molecular gas content traces the star formation activity in a galaxy. We also examine the dependence of H$_2$ and CO formation on dust evolution. Although the effects of hydrodynamical evolution of the ISM on H$_2$ and CO abundances are already investigated \citep[e.g.][]{Duarte-Cabral:2015aa,Richings:2016ab}, we are able to newly address the effects of dust evolution (both dust abundance and grain size distribution) on them and on the star formation law for the first time.

This paper is organized as follows. In Section~\ref{Model}, we review the dust evolution model we adopt and explain the calculation methods of H$_2$ and CO abundances. In Section~\ref{Results}, we present our calculation results. In Section~\ref{obs}, we compare our results with observed star formation laws. In Section~\ref{Discussion}, we discuss the dependence on some modelling features and effects of the dust evolution. In Section~\ref{Conclusions}, we summarize our work.

\section{Model}
\label{Model}
The main focus of this paper is the effect of dust evolution on the star formation law (especially the abundances of H$_2$ and CO). The dust evolution in a galaxy is represented by the simulation of an isolated disc galaxy in A17. The surface density of star formation rate is an output of the simulation, which assumes that star formation occurs in dense regions. The H$_2$ and CO surface densities are calculated in a consistent manner with the dust properties (dust abundance and grain size distribution) calculated in the simulation with a post-processing method developed in this paper. Note  that the star formation rate in the simulation is estimated by the gas density and is not related to molecular gas abundances. In other words, we neglect the possible effect of molecular gas abundance on star formation \citep{Hirashita:2002aa,Gnedin:2009aa,Krumholz:2012aa}; this serves to focus on the effect of dust evolution on the molecular gas content under a fixed star formation model. Moreover, whether or how much the molecular abundance affects the star formation in the metallicity range of interest in this paper (mainly $\gtrsim 0.1 Z_\odot$) is not clear. Thus, we completely `decouple' the molecular abundance from the star formation rate in the simulation. In this section, we first briefly review A17's simulation. After that, we explain how we calculate the H$_2$ and CO abundance.

\subsection{Galaxy simulation with dust evolution}
\label{dust}
In this paper, we adopt some snapshots of the hydrodynamic galaxy evolution simulation developed by A17. We give the time $t$ as measured from the beginning of the simulation. In their simulation, dust evolution is calculated consistently with the physical properties of the gas. We adopt a slightly modified version used in \citet{Hou:2017aa}. Below we briefly explain the simulation. We refer the interested reader to the above references for further details.

They simulated an isolated disc galaxy with a mass scale similar to the Milky Way using the modified version of \textsc{GADGET-3} $N$-body/SPH code \citep{Springel:2005aa}. The simulation includes dark matter, gas, and star particles. The Grackle\footnote{\href{https://grackle.readthedocs.org/}{https://grackle.readthedocs.org/}} chemistry and cooling library \citep{Bryan:2014aa,Kim:2014aa} are adopted to solve the non-equilibrium primordial chemistry network. The initial condition is taken from the low-resolution isolated galaxy model of \textsf{AGORA} simulations \citep{Kim:2014aa} with the minimum gravitational softening length of $\epsilon_\mathrm{grav}$ = 80 pc, the mass of gas particle $m_\mathrm{gas} = 8.59~\times~10^4$ M$_{\odot}$, and total gas mass of $8.59~\times~10^{9}$ M$_{\odot}$ for the entire galaxy. We summarize important parameters in Table \ref{tab:phys_quan_simu} and we refer the interested reader to table 1 in A17 for more details.
In this model, the local SFR density is estimated as 
\begin{equation}
\begin{aligned}
	\frac{d\rho_{*}}{dt} = \varepsilon_\mathrm{SF} \frac{\rho_\mathrm{gas}}{t_\mathrm{ff}},
	\label{eq:sfr1}
\end{aligned}
\end{equation}
where $\varepsilon_\mathrm{SF} = 0.01$ is the star formation efficiency, $\rho_\mathrm{gas}$ is the gas mass density and $t_\mathrm{ff}$ is the local free-fall time estimated as $t_\mathrm{ff} = \sqrt{{3\pi}/{32G\rho_\mathrm{gas}}}$ ($G$ is the gravitational constant). Star formation is allowed only in gas particles with $n_\mathrm{gas} > 10$ cm$^{-3}$, where $n_\mathrm{gas}$ is the gas number density. As noted above, this star formation recipe in the simulation is independent of the molecular abundance. If the star-forming regions with $n_\mathrm{gas} > 10$ cm$^{-3}$ are fully molecular, the star formation law investigated in this paper (i.e. the relation between the surface densities of molecular gas mass and SFR) reflects the above star formation recipe. The star particles are stochastically created from gas particles as described in \citet{Springel:2003aa} consistently with the above SFR.
Early stellar feedback from stellar winds and radiation of massive stars and SN feedback are also considered (see Section 2.1 of A17 for detailed references). The feedback and metal enrichment from stars are assumed to occur almost instantaneously (in 4 Myr) after the star formation.
The star formation rate of the entire galaxy is almost constant ($\sim 1$ M$_{\odot}$ yr$^{-1}$) throughout the evolution. The gas fraction at $t=10$ Gyr ($\sim 0.1$) is similar to that of the Milky Way.

\begin{table}
	\centering
	\caption{Initial physical quantities in our simulated galaxy that are important for this work (H$_2$ formation and CO formation). We allow the gas smoothing length to be 10 per cent of the gravitational softening length, so the final gas smoothing length reaches about 30 pc in the highest density regions. We refer the interested reader to table 1 in A17 for details.}
	\label{tab:phys_quan_simu}
	\begin{tabular}{lr} 
		\hline
		Physical parameter & Value\\
		\hline
		Total gas mass & $8.59~\times~10^{9}~$ M$_\odot$\\
		Total dark matter mass & $1.25~\times~10^{12}$ M$_\odot$\\
		Total mass & $1.3~\times~10^{12}~$M$_\odot$\\
		Gas particle mass & $8.59~\times~10^{4}~$M$_\odot$\\
		Dark matter particles mass & $1.25~\times~10^{7}$ M$_\odot$\\
		Grav. softening length & 80 pc\\
		\hline
	\end{tabular}
\end{table}

For dust evolution, the two-size dust model developed by \citet{Hirashita:2015aa} is adopted instead of a continuous grain size distribution because of the limitation in the computational cost. In this model, the whole grain size range is divided into two at $a\sim 0.03~\micron$, based on a previous calculation of full grain size distributions by \citet{Asano:2013aa}, and the evolution of the large and small abundances is considered. For each gas particle, the evolution of the dust-to-gas ratio is calculated. The dust-to-gas ratio on a gas particle is defined for each grain size and each species as
\begin{equation}
\begin{aligned}
	\mathcal{D}_{i,\mathrm{X}} \equiv m_{i,\mathrm{X}} / m_\mathrm{gas},
	\label{eq:d2g}
\end{aligned}
\end{equation}
where $m_{i,\mathrm{X}}$ is the mass of the relevant dust size and species with $i$ indicating small or large dust grains ($i$ = S or L), and X denoting the dust species (Si and C for silicate and carbonaceous dust, respectively), and $m_\mathrm{gas}$ is the total mass of the gas particle. We omit the label for each gas particle. 

Stars produce dust at their final stage of evolution. Because the simulation assumes almost instantaneous metal ejection after star formation, we only consider dust formation by SNe and neglect the contribution from asymptotic giant branch (AGB) stars. In fact, this assumption does not affect our results significantly, because in our simulated galaxy, the major source of dust is dust growth by accretion at the ages \citep[$\gtrsim 1$ Gyr;][]{Valiante:2009aa} when AGB stars would contribute to the dust abundance \citep{Hou:2017aa}. It is also assumed that SNe produce only large grains \citep{Nozawa:2007aa,Bianchi:2007aa,Asano:2013aa}. For the dust destruction in the ISM by SN shocks, because the simulation does not resolve individual SNe, an analytic subgrid treatment is applied. In addition, the following dust processing mechanisms in the ISM are considered: shattering, coagulation, and accretion. Shattering is a process in which large grains collide with each other and fragment into small grains. Because shattering only occurs in the diffuse ISM \citep{Yan:2004aa, Hirashita:2009aa}, shattering is considered only in gas particles with $n_\mathrm{gas}~<$ 1 cm$^{-3}$. Two grain growth mechanisms, coagulation and accretion are considered. Coagulation is a process in which small grains collide with each other and form large grains, while accretion is a process in which small grains gain mass by accreting gas-phase metals. Both coagulation and accretion are only allowed in dense clouds, which are not resolvable in the simulation. Thus, subgrid models of those two processes are applied to gas particles with $n_\mathrm{gas}~>$ 10 cm$^{-3}$ and $T_\mathrm{gas}<10^{3}$ K ($T_\mathrm{gas}$ is the gas temperature) by assuming that mass fraction $f_\mathrm{dense}$ of those gas particles are dense clouds with a fixed temperature ($T_\mathrm{dense}$) and a fixed gas number density ($n_\mathrm{dense}$). We adopt $f_\mathrm{dense} = 0.5$, $T_\mathrm{dense} = 50$ K, and $n_\mathrm{dense} = 1000$ cm$^{-3}$ (A17).

For each gas particle, the time evolution of dust-to-gas ratio from time $t$ to the next time step $t + \Delta t$ is, therefore, formulated as the following equations:
\begin{equation}
\begin{aligned}
	\mathcal{D}_\mathrm{L,X}(t + \Delta t) & = \mathcal{D}_\mathrm{L,X}(t) - \Delta \mathcal{D}^\mathrm{SN}_\mathrm{L,X}(t) \\
	& +  \left ( \frac{\mathcal{D}_\mathrm{S,X}(t)}{\tau_\mathrm{co}} - \frac{\mathcal{D}_\mathrm{L,X}(t)}{\tau_\mathrm{sh}} \right ) \Delta t\\
	& + f_\mathrm{in,X}\mathcal{Y}^{\prime}_\mathrm{X}\frac{\Delta M_\mathrm{return}}{m_\mathrm{gas}} \left (1 - \delta \right),  \\
	\label{eq:dust_large}
\end{aligned}
\end{equation}
\begin{equation}
\begin{aligned}
	\mathcal{D}_\mathrm{S,X}(t + \Delta t) & = \mathcal{D}_\mathrm{S,X}(t) - \Delta \mathcal{D}^\mathrm{SN}_\mathrm{S,X}(t) \\
	 & + \left( \frac{\mathcal{D}_\mathrm{L,X}(t)}{\tau_\mathrm{sh}} - \frac{\mathcal{D}_\mathrm{S,X}(t)}{\tau_\mathrm{co}} + \frac{\mathcal{D}_\mathrm{S,X}(t)}{\tau_\mathrm{acc}} \right) \Delta t, \\
	\label{eq:dust_small}
\end{aligned}
\end{equation}
where $\Delta \mathcal{D}^\mathrm{SN}_{i,\mathrm{X}}(t)$ is the decrease of dust-to-gas ratio by SN destruction in the ISM, $f_\mathrm{in,X}$ is the dust condensation efficiency in stellar ejecta, $\Delta M_\mathrm{return}$ is the returned gas mass from stars, $\delta$ is the fraction of the newly formed dust destroyed by SN shocks, $\mathcal{Y}^{\prime}_\mathrm{X}$ is the metal yield, and $\tau_\mathrm{sh}, \tau_\mathrm{co},$ and $\tau_\mathrm{acc}$ are the time-scales of shattering, coagulation, and accretion, respectively. For the metal yield, they adopted a total metal yield of $\mathcal{Y}^{\prime}_{Z} = 0.02896$ and derived the Si and C yields by assuming the solar abundance pattern \citep{Lodders:2003aa}. The detailed treatment of separate dust species, however, does not affect the results in this paper, since we only use the total dust abundance in our calculations of molecular abundances. The time-scales, $\tau_\mathrm{sh}, \tau_\mathrm{co}$, and $\tau_\mathrm{acc}$, are estimated by collision time-scales of relevant species and are dependent on the metallicity and/or the dust-to-gas ratio \citep[see][for their expressions]{Hou:2017aa}. We assume that Si occupies a fraction of 0.166 in silicate while C is the only constituent of carbonaceous dust. We adopt solar metallicity $Z_{\odot} = 0.02$ following A17. The solar metallicity is used only for the normalization of the resulting metallicity and does not affect the dust evolution calculations. The initial metallicity and dust-to-gas ratios are assumed to be zero.

The dust evolution obtained by A17 is summarized as follows. At early stages ($t \lesssim$ 0.3 Gyr), dust-rich regions are limited to the central region of the galaxy because of the high star formation activities there. In this epoch, the abundance ratio of small grains to large grains is small because the dust abundance is dominated by stellar dust production. As the galaxy evolves, dust is distributed more widely and the small grain abundance is enhanced because of shattering and accretion. At later stages ($t \gtrsim$ 1 Gyr), accretion further enhances the small-grain abundance in the entire galactic regions. In this epoch, the large-grain abundance also increases because of coagulation, especially in the central part where the gas density is high.

In this paper, we neglect the difference between silicate and carbonaceous dust because (i) the difference in the H$_2$ formation rate between these species is not well known, and (ii) the abundance ratio between silicate and carbonaceous dust has uncertainty as discussed in \citet{Hou:2017aa}. Thus, we only use the dust-to-gas ratio $\mathcal{D}_{i} \equiv \sum_{\mathrm{X}} \mathcal{D}_{i,\mathrm{X}}$, where we still distinguish between $i$ = S and L.
This treatment, however, underestimates the total dust amount by a factor of $\sim 2$ compared with A17. Since the chemical evolution model has a factor 2 uncertainty in the metal yield, we use \citet{Hou:2017aa} keeping in mind a possible underestimate of dust-to-gas ratio.

\subsection{H$_2$ abundance}
\label{H2ab} 
We calculate the H$_2$ abundance of each gas particle using the relevant quantities calculated in the simulation reviewed in Section~\ref{dust}. Although there have been detailed analytic models for the H$_2$ abundance \citep[][]{Hirashita:2005aa,Krumholz:2008aa,Krumholz:2009aa,McKee:2010aa}, we newly consider the effect of grain size in this paper. The H$_2$ abundance is quantified by the molecular fraction $f_\mathrm{H_2}$ defined as
\begin{equation}
    f_\mathrm{H_2} \equiv \frac {2 n_\mathrm{H_2}} {n_\mathrm{HI} + 2 n_\mathrm{H_2}},
	\label{eq:fH2}
\end{equation}
where $n_\mathrm{H_2}$ and $n_\mathrm{HI}$ are the number densities of molecular hydrogen and atomic hydrogen, respectively. We neglect ionized hydrogen (H{\sc{ii}}) because we are interested in cold ($\leq 10^{4}$ K) gas in this paper. We assume the equilibrium between H$_{2}$ formation and destruction to obtain $f_\mathrm{H_2}$ as explained in what follows. We later comment on this equilibrium assumption.

First, we consider H$_2$ formation. We neglect H$_2$ formation in the gas phase, which is much less efficient than H$_2$ formation on grain surfaces if $Z \gtrsim 0.01 Z_{\odot}$ \citep{Todini:2001aa, Hirashita:2002aa}. Because we are only interested in regions with high H$_2$ abundance, neglecting inefficient gas-phase H$_2$ formation does not affect the discussion in this paper. The increasing rate of $f_\mathrm{H_2}$ by H$_2$ formation on dust is described as
\begin{equation}
	\left[ \frac{df_\mathrm{H_2}}{dt}  \right]_\mathrm{form} = 2 \sum_{i}^{} (1 - f_\mathrm{H_2}) R^{i}_\mathrm{H_2,dust} n_\mathrm{H},
	\label{eq:ifh2}
\end{equation}
where $n_\mathrm{H}$ is the number density of hydrogen nuclei, and $R^{i}_\mathrm{H_2,dust}$ (superscript $i$ indicates small or large grains) is the reaction rate coefficient. This rate coefficient is calculated by \citep[][HH17]{Yamasawa:2011aa}
\begin{equation}
	R^{i}_\mathrm{H_2, dust} =\frac{3 \mathcal{D}_i \mu m_\mathrm{H} S_\mathrm{H} \bar{v} \langle a^2 \rangle_i}{8 \pi s \langle a^3 \rangle_i},
	\label{eq:reaction_rate}
\end{equation}
where $\langle a^{n}\rangle_{i}$ ($n$ = 2 or 3) is the $n$th moment of the grain radius, $\mu$ is the gas mass per hydrogen nucleus (dimensionless), $m_\mathrm{H}$ is the atomic mass of hydrogen, $S_\mathrm{H}$ is the sticking coefficient, $\bar{v}$ is the thermal velocity and $s$ is the grain material density. We adopt $\mu$ = 1.4, $s$ = 3.3 g cm$^{-3}$, $\langle a^3 \rangle_\mathrm{S} / \langle a^2 \rangle_\mathrm{S} = 3.8 \times 10^{-3} \mu$m  and $\langle a^3 \rangle_\mathrm{L} / \langle a^2 \rangle_\mathrm{L} = 7.6 \times 10^{-3} \mu$m  (HH17).
The thermal velocity is estimated as \citep{Spitzer:1978aa}
\begin{equation}
\begin{aligned}
	\bar{v} =\sqrt\frac{8 k_\mathrm{B} T_\mathrm{gas}} {\pi m_\mathrm{H}},
	\label{eq:thermal_speed}
\end{aligned}
\end{equation}
where $k_\mathrm{B}$ is the Boltzmann constant.
The sticking coefficient is given by \citep{Hollenbach:1979aa,Omukai:2005aa}
\begin{equation}
\begin{aligned}
    S_\mathrm{H} = & \left[ 1 + 0.04 \left( T_\mathrm{gas} + T_\mathrm{d} \right) ^{0.5} + 2 \times 10^{-3} T_\mathrm{gas} + 8 \times 10^{-6} T_\mathrm{gas}^2 \right] ^{-1} \\
          & \times \left\{  1 + \exp \left[ 7.5 \times 10^2 \left( \frac{1}{75} - \frac{1}{T_\mathrm{d}} \right) \right] \right\} ^{-1} ,\\
	\label{eq:sticking_co}
\end{aligned}
\end{equation}
where $T_\mathrm{d}$ is the dust temperature. The dust temperature is estimated as \citep{Hirashita:2005aa} 
\begin{equation}
    T_\mathrm{d} = 15 \chi^{\frac{1}{6}} \left( \frac{a}{0.1 \mu \mathrm{m}} \right) ^{-\frac{1}{6}} \mathrm{K},
	\label{eq:dust_t}
\end{equation}
where $\chi$ is the UV interstellar radiation field (ISRF) normalized to the solar neighbourhood value given by \citet{Habing:1968aa}. 

Now we explain how to evaluate $\chi$. Since the dissociating radiation is dominated by recently formed stars, the ISRF is related to the surface density of SFR ($\overline{\Sigma}_\mathrm{SFR}$) as \citep{Hirashita:2003aa},   
\begin{equation}
   \overline{\chi} = 8.3 \times 10^{2}\left(
   \frac{\overline{\Sigma}_\mathrm{SFR}}{\mathrm{M_{\odot}}~\mathrm{yr}^{-1}~\mathrm{kpc}^{-2}}\right),
	\label{eq:sfrchi}
\end{equation}
where $\overline{\Sigma}_\mathrm{SFR}$ is the SFR surface density averaged for nearby gas particles with appropriate weighting. Here, we adopted the \citet{Chabrier:2003aa}  initial mass function (IMF) instead of the \citet{Salpeter:1955aa} IMF used in \citet{Hirashita:2003aa}. To calculate $\overline{\Sigma}_\mathrm{SFR}$, we consider the contribution from nearby gas particles where the SFR of each gas particle is given by the simulation (equation \ref{eq:sfr1}). We divide the disc into two-dimensional grids and adopt a grid size of $L_\mathrm{grid} = 200$ pc, giving an area of $A_\mathrm{grid} = 4 \times 10^4$ pc$^2$. This grid size remains the same unless otherwise stated. The radiation from nearby grids contributes to the ISRF. Using equation~(\ref{eq:sfrchi}), $\overline{\chi}$ at a certain grid point ($l_0, m_0$) is estimated as
\begin{equation}
	 \overline{\chi}(l_0,m_0) = \frac{\xi}{A_\mathrm{grid}} \frac{\sum_{d<d_0} \mathrm{SFR}(l_1,m_1)W(l_0,m_0;l_1,m_1)}{\sum_{d<d_0} W(l_0,m_0;l_1,m_1)},\\
	\label{eq:sigmasfr}
\end{equation}
where $\xi =8.3\times 10^2/(\mathrm{M_{\odot}}~\mathrm{yr}^{-1}~\mathrm{kpc}^{-2})$
(i.e.\ $\overline{\chi}=\xi\overline{\Sigma}_\mathrm{SFR}$; see equation \ref{eq:sfrchi}), $W(l_0,m_0;l_1,m_1)$ is the weighting of SFR($l_1,m_1$) at grid point ($l_1, m_1$) and we adopt $d_0 = 3$ kpc to take into account only local SFRs. To reduce computational time, $W(l_0,m_0;l_1,m_1)$ is simply determined by the grid in which the gas particle is located  instead of the actual coordinate. The weight of each gas particle is inversely proportional to the square of the distance based on the scaling of radiation flux. The weighting of the contribution from grid $(l_1,m_1)$ to grid $(l_0,m_0)$ is estimated as
\begin{equation}
	W(l_0,m_0;l_1,m_1) = \frac{1}{(l_0-l_1)^2 + (m_0-m_1)^2}.
	\label{eq:wk}
\end{equation}
We adopt $W = 4$ for gas particles located inside grid$(l_0,m_0)$. We calculate the interstellar UV radiation field $\chi$ using equations~(\ref{eq:sfrchi})--(\ref{eq:wk}).

Next, we consider H$_2$ destruction by photodissociation. The changing rate of $f_\mathrm{H_2}$ by photodissociation is estimated as
\begin{equation}
	\left[ \frac{df_\mathrm{H_2}}{dt}  \right]_\mathrm{diss} = - R_\mathrm{diss} f_\mathrm{H_2},
	\label{eq:mfh2}
\end{equation}
where $R_\mathrm{diss}$ is the dissociation rate coefficient. This rate coefficient is given by \citep{Hirashita:2005aa}  
\begin{equation}
    R_\mathrm{diss} = 4.4 \times 10^{-11} \chi S_\mathrm{shield, H_2} S_\mathrm{shield, dust}~\left[\mathrm{s}^{-1}\right], \\
	\label{eq:photodissociation_rate}
\end{equation}
where $S_\mathrm{shield, H_2}~(\leq 1)$ and $S_\mathrm{shield, dust}~(\leq 1)$ are the correction factors for self-shielding and dust shielding, respectively.
The self-shielding factor is given by \citet{Draine:1996aa} as
\begin{equation}
\begin{aligned}
    S_\mathrm{shield, H_{2}} = \mathrm{min} \left[ 1, \left( \frac{\frac{1}{2} f_\mathrm{H_{2}} N_\mathrm{H}}{10^{14} \mathrm{cm}^{-2}} \right) ^{-0.75} \right],
	\label{eq:H2_shielid}
\end{aligned}
\end{equation}
where $N_\mathrm{H}$ is the column density of hydrogen nuclei.
The dust shielding factor is given by
\begin{equation}
    {S_\mathrm{shield, dust}} = \exp \left( - \sum_{i}^{} \tau_{\mathrm{LW},i} \right) ,
	\label{eq:dust_shieliding_factor}
\end{equation}
where $\tau_{\mathrm{LW},i}$ is the optical depth of dust component $i$ ($i = $ S or L) at the LW band. The dust optical depth is estimated as
\begin{equation}
    \tau_\mathrm{LW, i} = \left( \frac {\tau_\mathrm{LW}} {N_\mathrm{H}} \right) _{0.01,i} \left( \frac{\mathcal{D}_{i}}{0.01} \right) N_\mathrm{H} ,
	\label{eq:optical_depth}
\end{equation}
where $\left( \frac {\tau_\mathrm{LW}} {N_\mathrm{H}} \right) _{0.01,i}$ is the optical depth of dust component $i$ ($i = $ S or L) for $\mathcal{D}_{i}$ = 0.01 at the LW band, normalized to the hydrogen column density. We adopt $\left( \frac {\tau_\mathrm{LW}} {N_\mathrm{H}} \right) _{0.01,\mathrm{S}} = 8.2~\times~10^{-21}$ cm$^{2}$ and $\left( \frac {\tau_\mathrm{LW}} {N_\mathrm{H}} \right) _{0.01,\mathrm{L}} = 2.3~\times~10^{-21}$ cm$^{2}$ \citep{Hirashita:2017aa}.

To obtain the equilibrium abundance of H$_2$, we equate the formation and destruction as 
\begin{equation}
	\left[ \frac{df_\mathrm{H_2}}{dt}  \right]_\mathrm{form} = -\left[ \frac{df_\mathrm{H_2}}{dt}  \right]_\mathrm{diss},
	\label{eq:equil2}
\end{equation}
and solve it for $f_\mathrm{H_2}$ using equations (\ref{eq:ifh2}) and (\ref{eq:mfh2}).

Our galaxy-scale simulation has difficulty in resolving individual molecular clouds, although it enables us to identify dense regions which potentially include molecular clouds. Therefore, we need an analytic model for the substructure of a dense region to treat molecular clouds. We hereby implement a subgrid model for the calculation of H$_2$ abundance. The dust evolution simulation we adopt (A17) also used a subgrid model for dust growth in dense regions. We adopt the same subgrid density structure as assumed in the simulation for consistency; that is, gas particles with $n_\mathrm{gas}~>~10$ cm$^{-3}$ and $T_\mathrm{gas} < 10^{3}$ K are identified as dense gas particles, and are assumed to contain dense clouds with a mass fraction of $f_\mathrm{dense}$ (Section \ref{dust}). We assume the temperature inside the dense clouds $T_\mathrm{dense} = 50$ K and the number density of hydrogen $n_\mathrm{H,d} = 10^{3}$ cm$^{-3}$, while for the rest (mass fraction $1-f_\mathrm{dense}$), which is referred to as the diffuse regions, the density is diluted to $(1-f_\mathrm{dense})$ times the density calculated in the simulation. The temperature in the diffuse regions is the same as the one calculated in the simulation. We adopt $f_\mathrm{dense} = 0.5$ following A17, unless otherwise stated.\footnote{Strictly speaking, if adopt values other than $f_\mathrm{dense}=0.5$ and $n_\mathrm{H,d}=10^3$ cm$^{-3}$ in our post-processing, the consistency with the simulation could be be lost. However, in reality, there is no guarantee that the molecular-rich regions exactly coincides with the dust-growing regions. We will adopt different values of $f_\mathrm{dense}$ and $n_\mathrm{H,d}$ in Section \ref{Discussion}.} Since both the dense and diffuse regions contribute to the column density, we calculate $f_\mathrm{H_2, dense}$ (the molecular fraction inside the dense clouds) and $f_\mathrm{H_2,diff}$ (the molecular fraction in the diffuse region) using equation~(\ref{eq:equil2}) under the common column density $N_\mathrm{H}$, which is estimated as $N_\mathrm{H} = n_\mathrm{H}h$, where and $h$ is the smoothing radius. Because the dense clouds are not resolved in the simulation, $h$ is just used for the effective length of the dense gas particle. This treatment neglects the density structure below a scale of $h$. Some simulations utilize the density gradient to obtain an appropriate length scale \citep{Gnedin:2009aa}, but this does not provide a stable estimate of the length scale in our simulation because of the poor spatial resolution. We also confirmed that the hydrogen column density given by our method is typically $\sim 10^{22}$ cm$^{-2}$, which is consistent with the typical column density of molecular clouds in the Milky Way \citep{Schneider:2016aa}.
We further tested our results against a higher resolution run (10-pc resolution) at $t=1$ Gyr, and confirmed that the H$_2$ abundance is insensitive to the resolution. The CO abundance is not sensitive to the resolution in the solar-metallicity environment except at the low surface density end of the star formation law, where the low resolution run tends to overproduce the CO abundance by a factor of 2--3. The CO abundance is underestimated in the low resolution also at low metallicity $Z\lesssim 0.4$ Z$_{\sun}$; however, CO is difficult to detect at such a low metallicity, so that it is difficult to test our prediction against observations. We address the importance of high spatial resolution for the CO abundance again in Section \ref{RZdep}.
Finally, the mean molecular fraction $\overline{f}_\mathrm{H_2}$ for each dense gas particle is obtained as
\begin{equation}
    \overline{f}_\mathrm{H_{2}} = f_\mathrm{dense} f_\mathrm{H_{2}, dense} + (1-f_\mathrm{dense}) f_\mathrm{H_{2}, diff}.
	\label{eq:meanfH2}
\end{equation}
We hereafter denote $\overline{f}_\mathrm{H_{2}}$ as $f_\mathrm{H_2}$ simply. For gas particles with $n_\mathrm{gas} < 10$ cm$^{-3}$ or $T_\mathrm{gas} > 10^{3}$ K, we use equation~(\ref{eq:equil2}) under the gas density and temperature given in the simulation together with $N_\mathrm{H} = n_\mathrm{H} h$ to obtain $f_\mathrm{H_2}$. 

As noted in HH17, the assumption of chemical equilibrium for the H$_2$ abundance is reasonable if we are interested in regions where the dust-to-gas ratio is not significantly lower than the Milky Way value. In such an environment, the time-scale of gravitational collapse is longer than the time-scale of the relevant chemical reactions. However, if the dust-to-gas ratio is much lower than the Milky Way value, non-equilibrium treatment could be desirable, especially for small-scale structures \citep{Gibson:2016aa,Hu:2016aa,Pallottini:2017aa}. The argument of non-equilibrium depends strongly on the lifetime or the sustaining mechanism of dense clouds. Because of the lack of spatial resolution, we simply assume equilibrium for the H$_2$ abundance in this paper. 

\subsection{CO abundance}
\label{COform}
For CO formation and destruction, directly solving chemical reactions is time consuming; therefore, we follow the method used by HH17 and originally developed by \citet{Feldmann:2012aa}. They utilized the CO abundance data already calculated for various physical conditions by \citet{Glover:2011aa} (hereafter GM11). CO forms in more embedded regions than H$_2$ because its self-shielding is weaker. However, our simulation is not capable of resolving such dense shielded regions. Thus, we apply a subgrid model to the calculation of CO abundance, following the above treatment for H$_2$. We assume that CO only forms inside gas particles with $n_\mathrm{gas} > 10$ cm$^{-3}$ and $T_\mathrm{gas} < 10^{3}$ K and that CO formation occurs in the subgrid dense clouds (Section~\ref{H2ab}) with $T_\mathrm{dense} = 50$ K, $n_\mathrm{H,d} = 10^3$ cm$^{-3}$, and $N_{\mathrm{H}} = n_{\mathrm{H}} h$. Thus, we neglect the possible dependence on density and temperature \citep{Narayanan:2012aa} but focus on the consistency with the dust evolution. Since we apply the same subgrid model over the entire regions, we neglect a possible environmental dependence of molecular cloud properties \citep{Pan:2017aa}. Although our treatment is rough, we emphasize that HH17 successfully explained the metallicity dependence of $X_\mathrm{CO}$ under those temperature and density. Thus, under the lack of resolution, HH17's analytic method will give us a useful guide for populating CO in the simulated galaxy. We briefly review their method in what follows, and refer the interested reader to HH17 for further details.

We define the CO abundance $f_\mathrm{CO}$ as the number ratio of CO molecules to hydrogen nuclei.\footnote{HH17 used the notation $x_\mathrm{CO}$ for $f_\mathrm{CO}$.} We assume that the CO abundance is determined by the dust extinction and the ISRF under a given metallicity; thus, we write the CO abundance as $f_\mathrm{CO} = f_\mathrm{CO}(A_V, \chi, Z)$. We impose an upper limit for $f_\mathrm{CO}$ by the carbon abundance $f_\mathrm{C} = 1.41 \times 10^{-4} Z/Z_{\odot}$. The basic idea of \citet{Feldmann:2012aa} and HH17 is to find a value of extinction $A'_V$ in the system of GM11 that satisfies $f'_\mathrm{CO}(A'_V,\,\chi '=1.7,\, Z)=f_\mathrm{CO}(A_V,\,\chi,\,Z)$ (GM11 calculated CO fraction $f'_\mathrm{CO}$ with $\chi'=1.7$ based on the Galactic radiation field derived by \citet{Draine:1978aa}), where the notations with a prime indicate the values in the system of GM11 and the notations without prime indicate the values from our simulation.

Before writing the main equation, we prepare the following fitting formulae based on GM11's results in order to write $f'_\mathrm{H_2}$ and $f'_\mathrm{CO}$ as a function of $A'_V$:
\begin{equation}
    f^{\prime}_\mathrm{H_{2}} = 1 - \exp(-0.45A^{\prime}_V),
	\label{eq:fH2prime}
\end{equation}
and 
\begin{equation}
    \log_{10} f^{\prime}_\mathrm{CO} = -7.64 + 3.89 \log_{10} A^{\prime}_V.
	\label{eq:xCOprime}
\end{equation}
The hydrogen column density in the GM11 system is related to $A^{\prime}_V$ as
\begin{equation}
    N^{\prime}_\mathrm{H} = \frac{A^{\prime}_V}{5.348 \times 10^{-22} (Z^{\prime} / Z_{\odot})~\mathrm{cm}^{2}}.
	\label{eq:NHprime}
\end{equation}
Note that we adopt $Z^{\prime} = Z$ so that the upper limits of CO abundance are the same in GM11's system and our system.

The basic assumption in \citet{Feldmann:2012aa} is that the CO formation rate is similar in the two systems if $f_\mathrm{CO}=f'_\mathrm{CO}$. Because the formation rate of CO is equal to the destruction (dissociation) rate of CO, the CO dissociation rate is similar in the two systems. Therefore, the problem of finding a condition that satisfies $f_\mathrm{CO}=f'_\mathrm{CO}$ reduces to the task of finding the same dissociation rate.
Since we are looking for a solution which satisfies $f_\mathrm{CO} = f^{\prime}_\mathrm{CO}$ (so $N_\mathrm{CO} = f^{\prime}_\mathrm{CO} N_\mathrm{H}$, where $N_\mathrm{CO}$ is the CO column density), the following equation approximately holds by the requirement that the dissociation rates in the two systems should be similar: 
\begin{equation}
\begin{aligned}
	& {\chi}S_\mathrm{dust}(A_{V\mathrm{,eff}})S_\mathrm{H_2}(f_\mathrm{H_2} N_\mathrm{H} / 2)S_\mathrm{CO}(f^{\prime}_\mathrm{CO}N_\mathrm{H})/N_\mathrm{H} \\
	& = 1.7S_\mathrm{dust}(A^{\prime}_V)S_\mathrm{H_{2}}(f^{\prime}_\mathrm{H_{2}} N^{\prime}_\mathrm{H} / 2)S_\mathrm{CO}(f^{\prime}_\mathrm{CO}N^{\prime}_\mathrm{H})/N^{\prime}_\mathrm{H},
	\label{eq:xCOre}
\end{aligned}
\end{equation}
where $S_\mathrm{dust}(A_{V,\mathrm{eff}})$, $S_\mathrm{H_2}(f_\mathrm{H_2} N_\mathrm{H} / 2)$, and $S_\mathrm{CO}(x^{\prime}_\mathrm{CO}N_\mathrm{H})$ are the shielding factors of CO-dissociating photons by dust, H$_2$, and CO, respectively, given by \citet{Lee:1996aa}; $f_\mathrm{H_2}$ and $N_\mathrm{H}$ are given by the calculations in Section~\ref{H2ab} and $A_{V,\mathrm{eff}} = A_{1000 \angstrom}/4.7$ is the effective $V$-band extinction converted from the extinction at 1000 $\angstrom$ to that at the $V$ band using the Milky Way extinction curve. Because $f^{\prime}_\mathrm{H_2}$, $f^{\prime}_\mathrm{CO}$, and $N^{\prime}_\mathrm{H}$ are functions of $A^{\prime}_V$ (equations~\ref{eq:fH2prime}--\ref{eq:NHprime}), the right-hand side of equation~(\ref{eq:xCOre}) is given as a function of $A^{\prime}_{V}$. Since $A_{V\mathrm{,eff}}$ represents dust shielding of CO-dissociating photons around 1000 $\angstrom$, $A_{1000 \angstrom} \simeq 1.086 \sum_{i}^{} \tau_{\mathrm{LW},i}$ which is calculated in equation~(\ref{eq:optical_depth}). In the end, equation~(\ref{eq:xCOre}) is reduced to an equation for $A^{\prime}_{V}$, which enables us to obtain $f_\mathrm{CO}$ (and $N_\mathrm{CO}=f_\mathrm{CO}N_\mathrm{H}$) through equation~(\ref{eq:xCOprime}) (recall that $f_\mathrm{CO} = f^{\prime}_\mathrm{CO}$).

Although the treatment of CO depends on the subgrid model, the simulation enables us to investigate the relation with star formation (i.e.\ star formation law). Moreover, the dissociation rate is estimated consistently with the local ISRF using the spatial distribution of star formation. Thus, we indeed predict how CO traces the star formation activity in various stages of galaxy evolution in a self-consistent manner with the star formation history and the metal/dust enrichment.

\subsection{CO-to-H$_2$ conversion factor}
\label{CO2H2}
Because H$_2$ is observationally traced by CO emission, it is important to clarify the relation between H$_2$ and CO. The key quantity of this relation is the CO-to-H$_2$ conversion factor defined as
\begin{equation}
	X_\mathrm{CO} = N_\mathrm{H_{2}}/W_\mathrm{CO},
	\label{eq:XCO}
\end{equation}
where $W_\mathrm{CO}$ is the intensity of the CO $J=1\to 0$ emission line.
We calculate $W_\mathrm{CO}$ using the following expression (GM11):
\begin{equation}
	W_\mathrm{CO} = T_{r} \Delta v \int^{\tau_{10}}_{0} 2 \beta (\tau) d\tau,
	\label{eq:WCO}
\end{equation}
where $T_\mathrm{r}$ is the observed radiation temperature (calculated later in equation~\ref{eq:Tr}), $\Delta v$ is the velocity width of the CO line, $\tau_{10}$ is the optical depth estimated in equation (\ref{eq:tau10}), and $\beta(\tau)$ is the photon escape probability as a function of the optical depth. The observed radiation temperature is given as
\begin{equation}
	T_\mathrm{r} = 5.5 \left( \frac{1}{e^{5.5/T_\mathrm{gas}} - 1} - \frac{1}{e^{5.5/T_\mathrm{CMB}} - 1} \right) ~\mathrm{K},
	\label{eq:Tr}
\end{equation}
where $T_\mathrm{CMB} = 2.73 (1+z)$ is the CMB temperature (we adopt $z$=0 in this paper, and we confirmed that the results are insensitive to the adopted CMB temperature). The photon escape probability is estimated as \citep{Tielens:2005aa}
\begin{equation}
	\beta (\tau) = 
		\begin{cases}  [1-\exp(-2.34\tau)]/(4.68\tau)~~~ \mathrm{if} ~ \tau \leq 7; \\
	                 \ 1/(4\tau[\ln(\tau/\sqrt{\pi})]^{1/2})~~~ \mathrm{if} ~ \tau > 7.
	    \end{cases}
	\label{eq:beta_tau}
\end{equation}
The optical depth $\tau_{10}$ is estimated as \citep{Tielens:2005aa,Feldmann:2012aa}
\begin{equation}
	\tau_{10} = 1.4 \times 10^{-16} (1 - e^{-5.5/T_\mathrm{gas}}) \left( \frac{\Delta v}{3~\mathrm{km~s^{-1}}} \right) ^{-1}  \left( \frac{N_\mathrm{CO}}{\mathrm{cm}^{-2}} \right).
	\label{eq:tau10}
\end{equation}
For the velocity width $\Delta v$, we adopt the typical value in \citet{Feldmann:2012aa}, $\Delta v = 3~\mathrm{km~s^{-1}}$, because the model is based on their model.

\subsection{Calculations of surface densities}
\label{grid}
Since we are interested in molecular-rich regions, we concentrate on the 20 kpc $\times$ 20 kpc region in the galactic disc. We divide the galactic disc into grids to calculate the distribution of surface densities for relevant quantities. We adopt the same grid size as in Section~\ref{H2ab} ($L_\mathrm{grid}=200$pc). In order to calculate surface densities, we use the kernel function to smooth the quantities of gas particles and divide the total contribution to the grid by the area of each grid $A_\mathrm{grid}$. We estimate the surface densities of H$_2$ and SFR denoted as $\Sigma_\mathrm{H_2}$ and $\Sigma_\mathrm{SFR}$, respectively. Note that $\Sigma_\mathrm{SFR}$ here is different from $\overline{\Sigma}_\mathrm{SFR}$, which is used to calculate the UV radiation field in Section~\ref{H2ab}. Since the mass fraction of C, O and other heavy elements in molecular clouds is usually small, we denote the surface density of total molecular gas as $\Sigma_\mathrm{mol} = 1.36\Sigma_\mathrm{H_2}$, taking helium into account. We also denote the CO-based total molecular gas surface density as $\Sigma^\mathrm{CO}_\mathrm{mol}$. We first calculate molecular gas mass based on the CO abundance and smooth the CO-based molecular gas mass of each gas particle. $M^\mathrm{CO}_\mathrm{mol}$ is estimated as
\begin{equation}
	M^\mathrm{CO}_\mathrm{mol} = 1.36 \frac{2 W_\mathrm{CO} X_\mathrm{CO,Gal}}{N_\mathrm{H}} M_\mathrm{H} ,
	\label{eq:sigmacoh2}
\end{equation}
where $M_\mathrm{H}$ is the total hydrogen mass, and $X_\mathrm{CO,Gal}$ is the Galactic conversion factor, $X_\mathrm{CO,Gal} = 2 \times 10^{20}$ cm$^{-2}$ (K km s$^{-1}$)$^{-1}$ \citep{Bolatto:2013aa}. $X_\mathrm{CO,Gal}$ is used in order to consistently compare our results with observations (note that $\Sigma^\mathrm{CO}_\mathrm{mol}$ equals to $\Sigma_\mathrm{mol}$ if the Galactic conversion factor is applicable).

Although we adopt $L_\mathrm{grid}=200$ pc, we also examined various grid sizes. Smaller grids give more detailed spatial distribution while larger grids average out the results. We confirmed that, as long as we are interested in the relations among the relevant surface densities, the results below are not significantly affected by the choice of the grid size. We also show the results with a different grid size $L_\mathrm{grid}=500$ pc in Section~\ref{obs} to compare with the observations. Since the sampling with $L_\mathrm{grid}=500$ pc is coarse, we adopt $L_\mathrm{grid}=200$ pc in Section~\ref{Results}. This small grid will be useful also for comparison with future high-resolution data.

\section{Results}
\label{Results}
\subsection{Molecular gas surface density distribution}
\label{RH2ab}
\begin{figure*}
	\begin{center}
	\includegraphics[width=2\columnwidth]{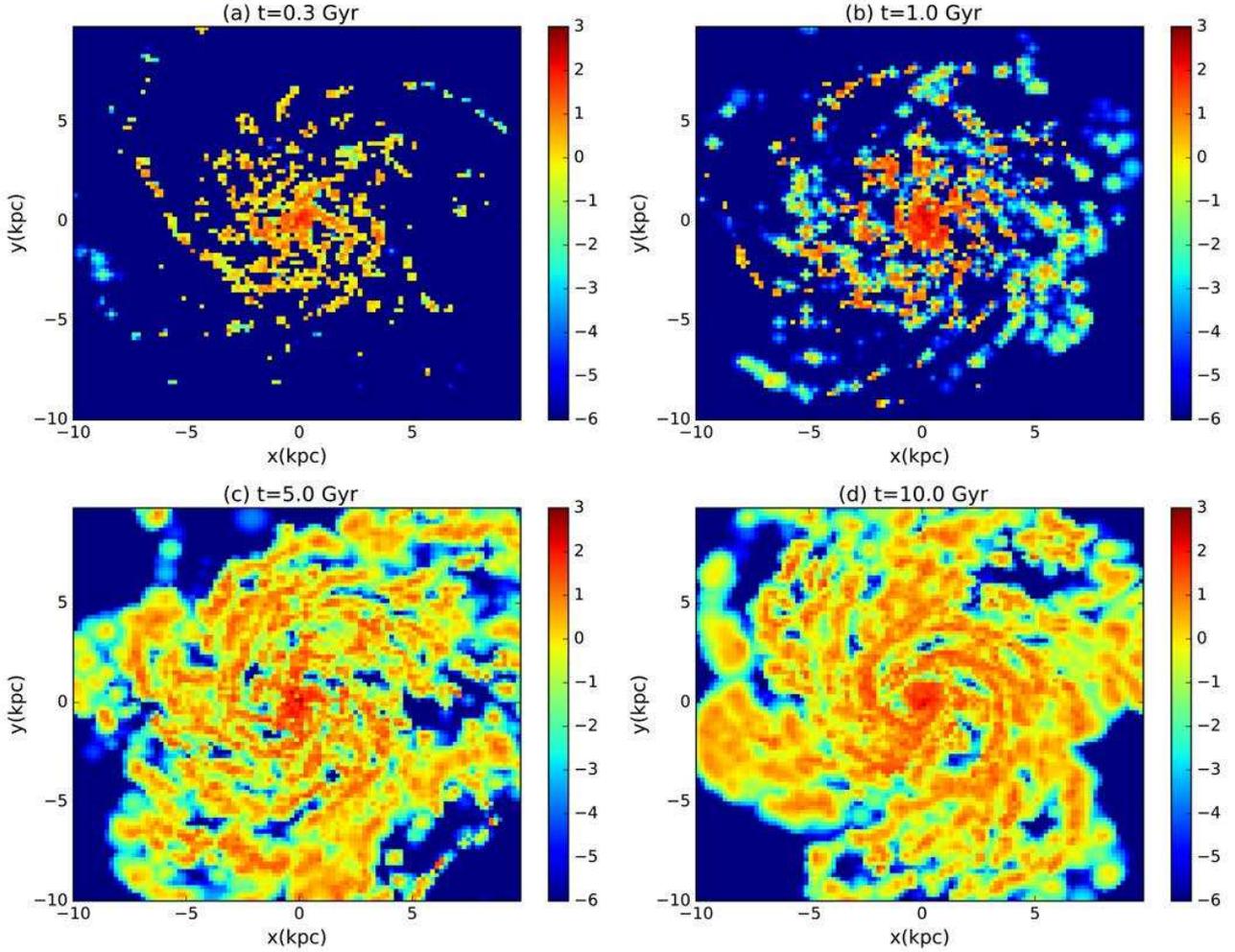}
	\caption{Distribution of molecular gas surface density ($\Sigma_\mathrm{mol}$) at $t=$ 0.3, 1, 5, and 10 Gyr in Panels (a), (b), (c) and (d), respectively. The colour indicates $\log_{10} \Sigma_\mathrm{mol}$ [M$_{\odot}$pc$^{-2}$], whose level is shown in the colour bar.}
	\label{fig:RH2xy}
	\end{center}
\end{figure*}

We present the surface density distribution of molecular gas at $t=$ 0.3, 1, 5, and 10 Gyr in Fig.~\ref{fig:RH2xy}. At $t=$ 0.3 Gyr, molecular-rich regions are limited to the spiral arms and have $\Sigma_\mathrm{mol} \sim 10^{-1} - 10^{2}$ M$_{\odot}$pc$^{-2}$ with the highest value in the central part of the galaxy. The star formation activity is the highest in the centre, so that dust and metals are produced most efficiently there. High star formation activities also produce strong radiation field, which destroys H$_2$ molecules; thus, H$_2$ is localized in well shielded dense regions. At $t=$ 1 Gyr, as the overall H$_2$ abundance is raised by dust enrichment, the distribution of molecular gas extends to the outer regions. The contrast between spiral arms and interarm regions is higher than at $t=$ 0.3 Gyr. This is because dust growth by accretion becomes efficient in dense regions as a result of metal enrichment. This trend is further pronounced at $t=$ 5 Gyr, when the molecular gas is distributed much more widely and the molecular gas surface density is higher than $10^{2}$ M$_{\odot}$pc$^{-2}$ in the central region. At $t=$ 5--10 Gyr, the molecular-rich regions are extended to the outer regions of the galactic disc and the molecular gas surface density reaches $10^{3}$ M$_{\odot}$pc$^{-2}$ in the central region. At this stage, the major source of dust in the entire disc is dust growth by accretion (A17).

\subsection{CO surface density distribution}
\label{RCOab}
\begin{figure*}
	\begin{center}
	\includegraphics[width=2\columnwidth]{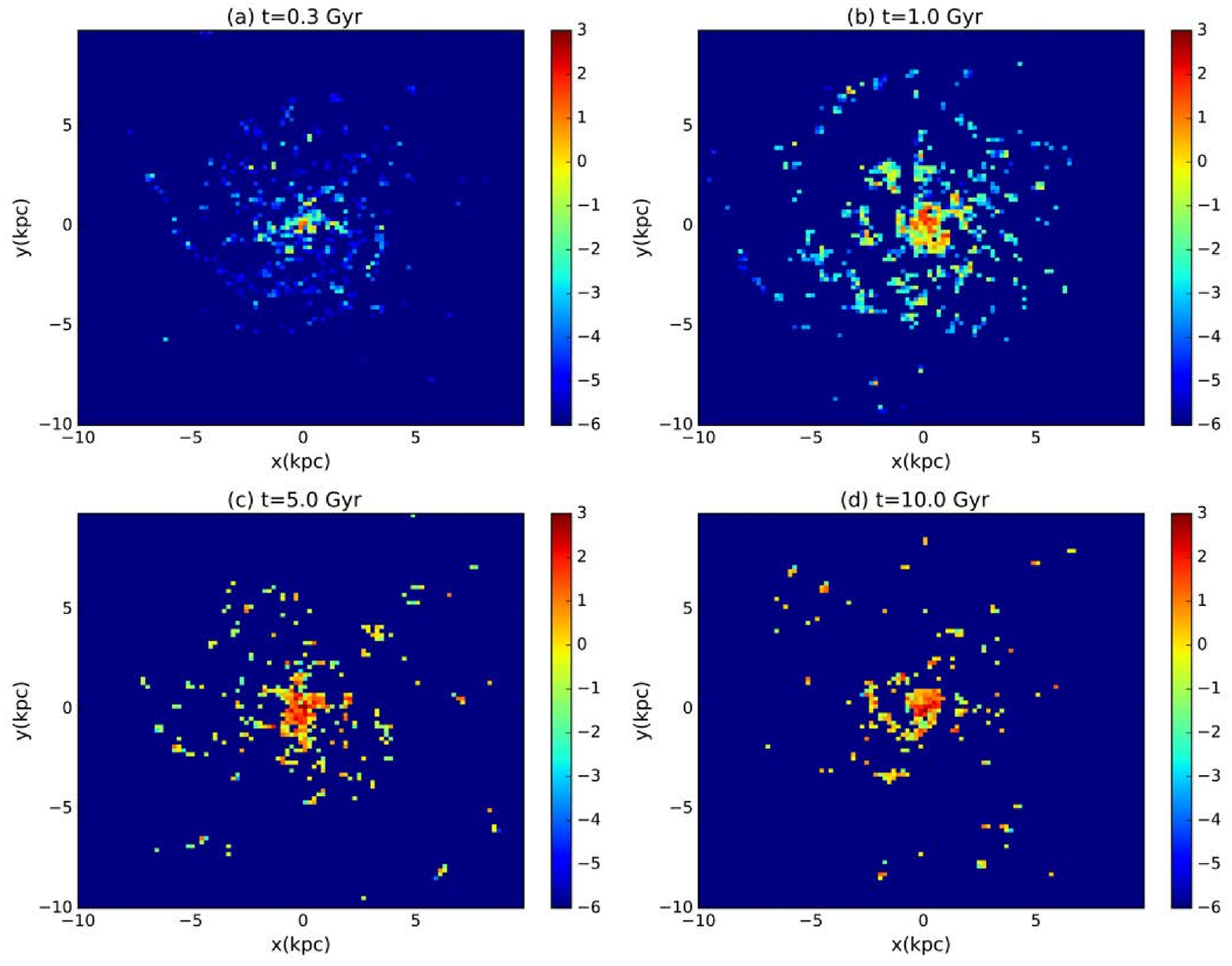}
	\caption{Same as Fig.~\ref{fig:RH2xy} but for $\Sigma^\mathrm{CO}_\mathrm{mol}$.}
	\label{fig:Rnew_H2xy}
	\end{center}
\end{figure*}

In Fig~\ref{fig:Rnew_H2xy}, we show $\Sigma^\mathrm{CO}_\mathrm{mol}$ (the CO-based molecular gas surface density; Section~\ref{grid}) at $t=$ 0.3, 1, 5, and 10 Gyr. At $t=$ 0.3 Gyr, the CO abundance is extremely low and an appreciable amount of CO is only found in a small part of the central region, where the dust abundance is high enough for CO formation. At $t=$ 1 Gyr, the CO abundance increases also in the outer dense regions. At $t=$ 5 and 10 Gyr, the abundance is much higher compared to younger ages but the distribution is still confined in `clumps' because the number of gas particles with $n_\mathrm{gas} > 10$ cm$^{-3}$ decreases after the significant gas consumption by star formation. At all ages, the distribution of CO-based molecular gas is more localized than that of H$_2$, since CO is populated only in dense regions by construction of our model (Section~\ref{COform}).

\subsection{Star formation law}
\label{RSFlaw}
\begin{figure*}
	\begin{center}
	\includegraphics[width=2\columnwidth]{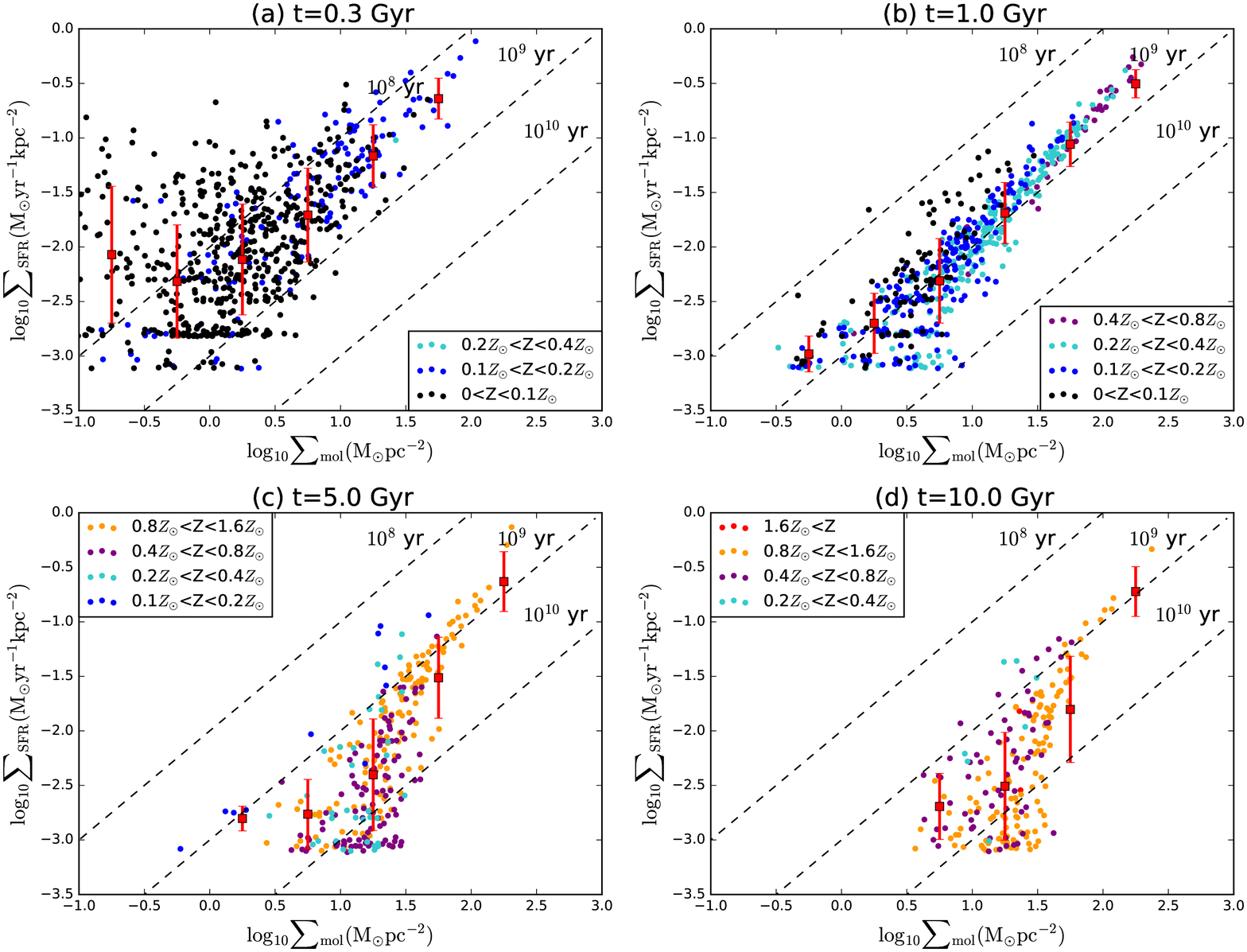} 
	\caption{Star formation rate surface density $\Sigma_{\mathrm{SFR}}$ from the simulation versus the calculated molecular gas surface density $\Sigma_\mathrm{mol}$ at $t=$ 0.3, 1, 5, and 10 Gyr in Panels (a), (b), (c) and (d), respectively. The colour of each point indicates the metallicity shown in the small window. Dotted lines indicate $\tau^\mathrm{mol}_\mathrm{dep} = 10^{8}$, $10^{9}$, and $10^{10}$ yr, from up to bottom. The running median with standard deviation are shown in red rectangles.}
	\label{fig:RSFRH2}
	\end{center}
\end{figure*}

Another important prediction in our work is the relation between the surface densities of SFR ($\Sigma_\mathrm{SFR}$) and molecular gas ($\Sigma_\mathrm{mol}$ or $\Sigma^\mathrm{CO}_\mathrm{mol}$), the so-called star formation law. We show this relation in Fig.~\ref{fig:RSFRH2} using the surface density of SFR ($\Sigma_\mathrm{SFR}$) and that of molecular gas ($\Sigma_\mathrm{mol}$). Each point represents each grid (an area of $4 \times 10^{4}$ pc$^{2}$, same as in Section~\ref{RH2ab} and \ref{RCOab}). In each panel in Fig.~\ref{fig:RSFRH2}, the mass-weighted mean metallicity in each grid is also shown by the colours. We also define the molecular gas depletion time as $\tau^\mathrm{mol}_\mathrm{dep} = \Sigma_\mathrm{mol} / \Sigma_\mathrm{SFR}$, which indicates the apparent time needed for star formation to use up molecular gas under the assumption that stars form from molecular gas. Since we do not assume that stars form from molecular gas (Section~\ref{dust}), $\tau^\mathrm{mol}_\mathrm{dep}$ is not necessarily the real gas depletion time, especially when H$_2$ formation is not efficient. If the dust abundance is high enough, H$_2$ traces the star-forming regions; in this situation, $\tau^\mathrm{mol}_\mathrm{dep}$ reflects the real depletion time. The constant molecular gas depletion time of $\tau^\mathrm{mol}_\mathrm{dep} = 10^{8}$, $10^{9}$, and $10^{10}$ yr are also plotted in Fig.~\ref{fig:RSFRH2}. The running median with a bin size of 0.5 dex for $\Sigma_\mathrm{mol}$ are shown with red squares with standard deviation.

At $t=$ 0.3 Gyr, the metallicities are lower than 0.4 $Z_{\odot}$; thus, H$_2$ formation is not efficient. The star formation law is scattered toward low molecular gas surface densities, and there is a trend that lower-metallicity points are more deviated. The molecular gas depletion time is less than $10^{9}$ yr because of the poor molecular gas content. At $t=$ 1 Gyr, as the metallicity increases to $\sim 0.8$ $Z_{\odot}$, the scatter becomes less than at $t=$ 0.3 Gyr, and a linear relation appears. The appearance of the linear relation can be interpreted as an establishment of the star formation law; in other words, the molecular gas content is strongly correlated with the star formation activity. At $t=$ 5--10 Gyr, most regions achieve solar metallicity, so that the scatter in the $\Sigma_\mathrm{SFR}$--$\Sigma_\mathrm{mol}$ relation decreases further. At all ages, there is a tendency that under similar $\Sigma_\mathrm{SFR}$, molecular gas abundance is higher for higher metallicity. This is due to an enhanced H$_2$ formation rate in an metal-rich (i.e. dust-rich) environment. The molecular gas depletion time is $\sim 10^9$ yr at $t=$ 1 Gyr and increases to $\sim$ a few $\times 10^9$ yr at $t=$ 5--10 Gyr as the molecular gas abundance becomes high.

\begin{figure*}
	\begin{center}
	\includegraphics[width=2\columnwidth]{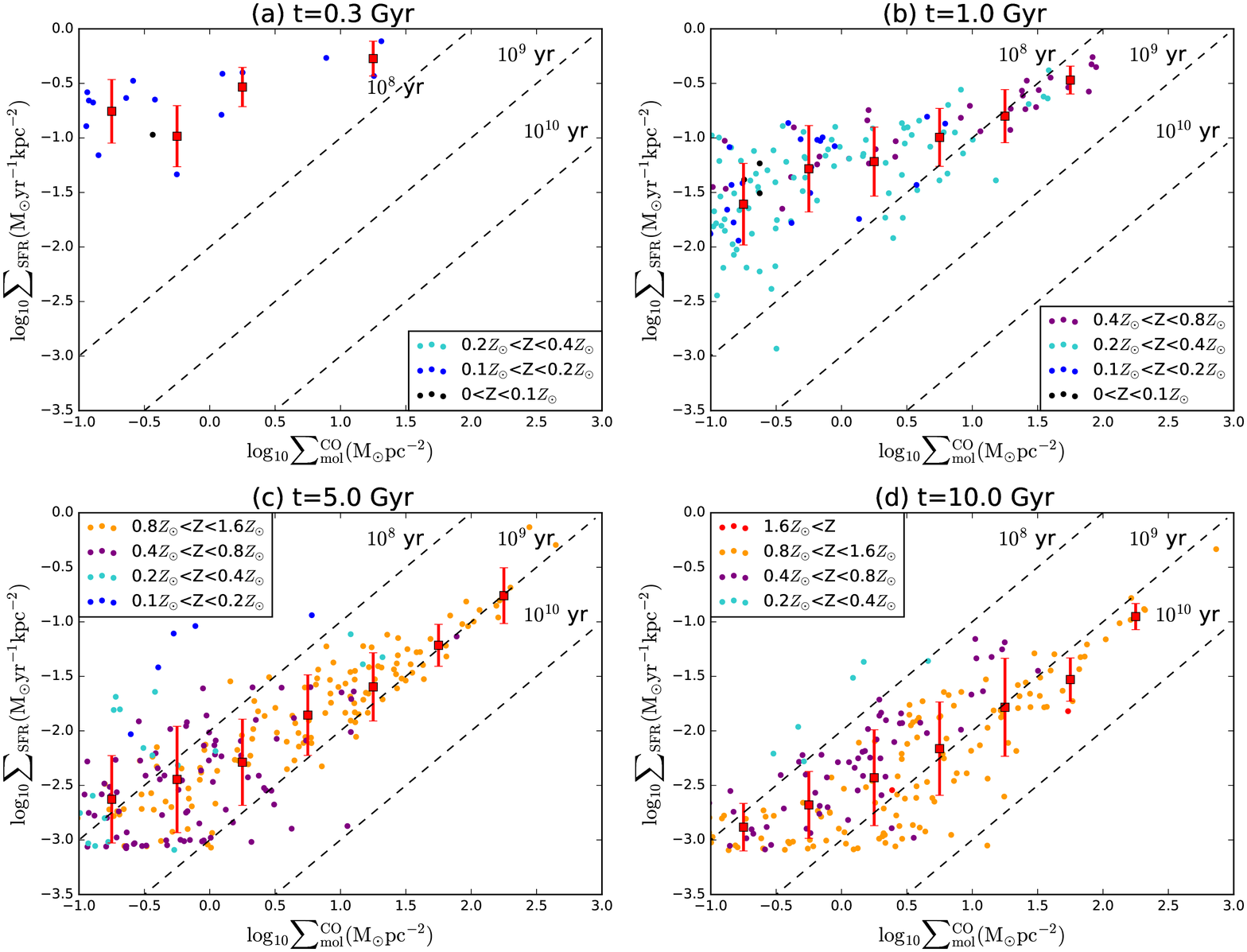}
	\caption{Same as Fig.~\ref{fig:RSFRH2} but for $\Sigma^\mathrm{CO}_\mathrm{mol}$ (the molecular gas surface density estimated from CO with the Galactic conversion factor).}
	\label{fig:RSFRnewH2}
	\end{center}
\end{figure*}

We here present the CO-based star formation law in Fig.~\ref{fig:RSFRnewH2}, where we adopt the CO-based molecular gas surface density converted by the Galactic conversion factor ($\Sigma^\mathrm{CO}_\mathrm{mol}$) (Section~\ref{grid}).  We show the metallicity of each data point in the same colour as in Fig.~\ref{fig:RSFRH2} and define the CO-based molecular gas depletion time as $\tau^\mathrm{CO}_\mathrm{dep} = \Sigma^\mathrm{CO}_\mathrm{mol} / \Sigma_\mathrm{SFR}$. This depletion time $\tau^\mathrm{CO}_\mathrm{dep}$ does not necessarily represent the real molecular gas depletion time because CO formation is confined to dense gas particles, and CO formation is inefficient at low-metallicity. The constant CO-based molecular gas depletion times are also plotted in Fig.~\ref{fig:RSFRnewH2}. In comparison with Fig.~\ref{fig:RSFRH2}, the distribution of data points shifts to the lower column density side. At $t=$ 0.3 Gyr, there are only a few data points because the low metallicity environments are unsuitable for CO formation. At $t=$ 1 Gyr, most of the data points are still scattered toward low $\Sigma^\mathrm{CO}_\mathrm{mol}$, in contrast to Fig.~\ref{fig:RSFRH2}b, in which a linear star formation law starts to be established. At $t=$ 5--10 Gyr, the data points are more scattered than in Figs.~\ref{fig:RSFRH2}c and d, but a linear star formation law is broadly established. There is a clear tendency that high metallicity regions show stronger CO emission.

\subsection{Metallicity dependence}
\label{RZdep}
\begin{figure*}
	\begin{center}
	\includegraphics[width=2\columnwidth]{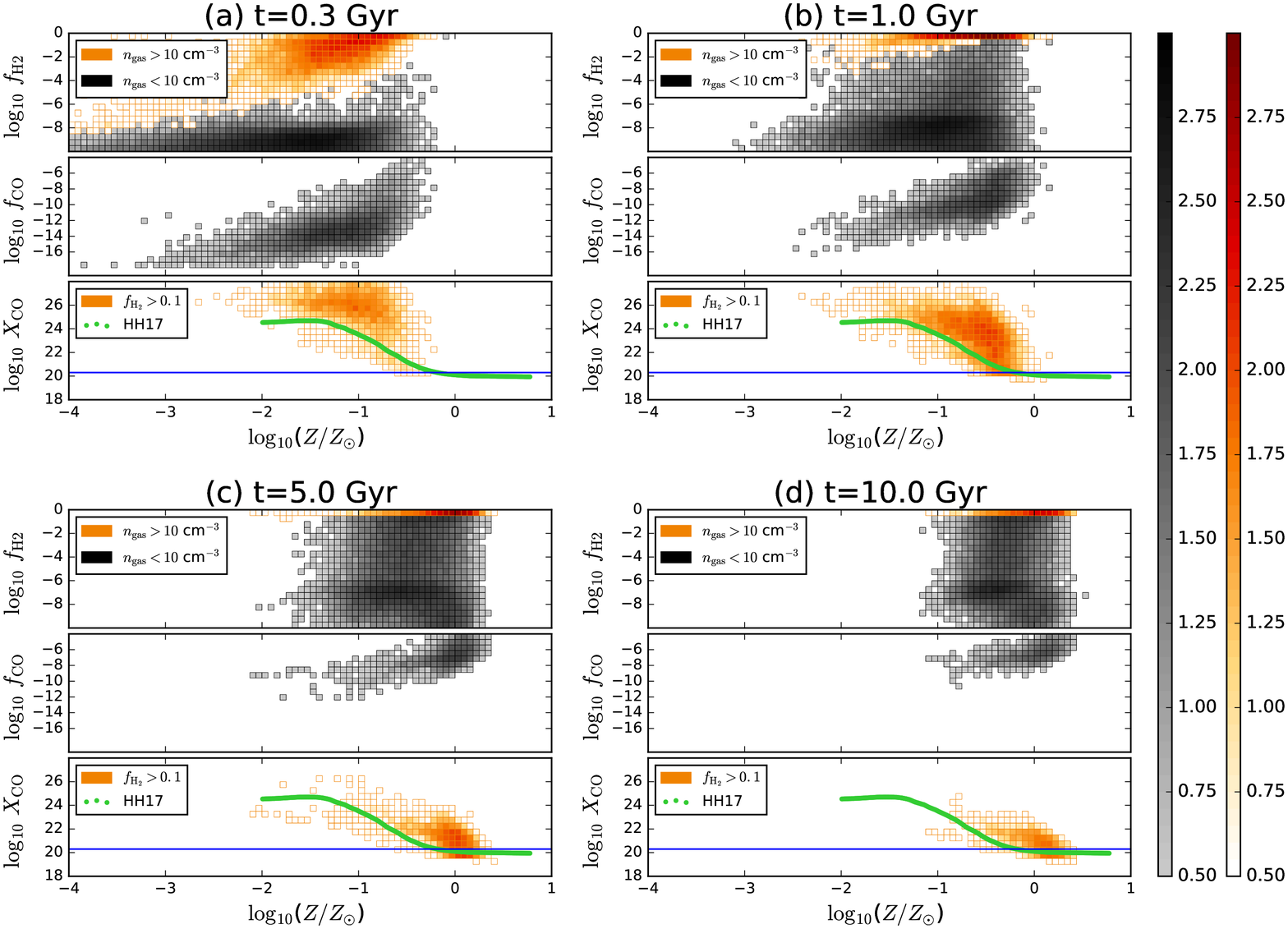}
	\caption{Molecular hydrogen fraction $f_\mathrm{H_2}$, CO fraction $f_\mathrm{CO}$, and CO-to-H$_2$ conversion factor $X_\mathrm{CO}$ as a function of metallicity at $t=$ 0.3, 1, 5, and 10 Gyr in Panels (a), (b), (c) and (d), respectively. The black and red colour scales represent the logarithmic number of data point in each grid on the diagram, as indicated in the colour bars on the right. In the $f_\mathrm{H_2}$ -- $Z$ relation, gas particles with $n_\mathrm{gas} > 10$ cm $^{-3}$ are shown in orange. In the $f_\mathrm{CO}$--$Z$ relation, only dense ($n_\mathrm{gas} > 10$ cm $^{-3}$) gas particles appear because we consider CO formation only in such dense gas particles. In the $X_\mathrm{CO}$ -- $Z$ relation, since we are interested in `molecular clouds', we only show gas particles with $f_\mathrm{H_2} > 0.1$ and the blue horizontal line shows the Milky Way conversion factor $X_\mathrm{CO,Gal} = 2 \times 10^{20}$ cm$^{-2}$ (K km s$^{-1}$)$^{-1}$. The theoretical prediction which reproduces the observational trend (HH17) is also shown (green solid curve).}
	\label{fig:R3Z}
	\end{center}
\end{figure*}

In Fig.~\ref{fig:R3Z}, we show the physical properties predicted by our model (molecular hydrogen fraction $f_\mathrm{H_2}$, CO fraction $f_\mathrm{CO}$, and CO-to-H$_2$ conversion factor $X_\mathrm{CO}$) as a function of metallicity in order to further examine the metallicity dependence discussed in the above subsections. Here we examine these relations for individual gas particles. To present the distribution of data points, we divide the diagram into grids and show the number of data points inside each grid. In the $f_\mathrm{H_2}$--$Z$ relation, to show the important role of gas density in H$_2$ formation, the distribution of gas particles with $n_\mathrm{gas} > 10$ cm$^{-3}$ are shown in orange. Recall that these dense gas particles are assumed to host dense clouds (Section \ref{H2ab}). In the $f_\mathrm{CO}$--$Z$ relation, since CO formation is only allowed in the dense clouds, only dense ($n_\mathrm{gas} > 10$ cm$^{-3}$) gas particles appear. In the $X_\mathrm{CO}$--$Z$ relation, since we are interested `molecular clouds', we only show gas particles with $f_\mathrm{H_2} > 0.1$, and the theoretical prediction by HH17 which reproduced the observational trend \citep{Arimoto:1996aa,Israel:1997aa,Bolatto:2008aa,Leroy:2011aa,Sandstrom:2013aa,Cormier:2014aa,Shi:2016aa} is also shown.

At $t=$ 0.3 Gyr, the data points are distributed over a wide metallicity range extending down to $\sim$ $10^{-4}~Z_{\odot}$. We observe that the dense gas particles are located on the upper branch of the $f_\mathrm{H_2}$--$Z$ relation because of self-shielding. There are less data points in the $f_\mathrm{CO}$--$Z$ and $X_\mathrm{CO}$--$Z$ relations because CO formation is only allowed in dense gas particles. In the $f_\mathrm{CO}$--$Z$ relation, we observe steep dependence of $f_\mathrm{CO}$ on $Z$. This is because UV shielding by H$_2$ is more efficient at high metallicity (HH17). In the $X_\mathrm{CO}$--$Z$ relation, we observe that gas particles with $f_\mathrm{H_2} > 0.1$ show a clear negative correlation between $X_\mathrm{CO}$ and $Z$, which is consistent with the observed trend \citep[e.g.][]{Bolatto:2013aa}. Most of these gas particles have $X_\mathrm{CO}$ larger than the theoretical prediction by HH17. This is because ISRF (or the dissociation rate) is high. Indeed the dense regions have $\chi \sim 10$--100 because they tend to be located near to the star-forming regions. 

At $t=$ 1 Gyr, particles with $Z<0.1$ Z$_{\odot}$ become rare. In the $f_\mathrm{H_2}$--$Z$ relation, gas particles with $n_\mathrm{gas}$ > 10 cm$^{-3}$ have high $f_\mathrm{H_2}$ and a significant fraction of them are fully molecular. Most of the particles with $n_\mathrm{gas}$ < 10 cm$^{-3}$ have low $f_\mathrm{H_2}$, but some of them achieve high $f_\mathrm{H_2} \sim 1$ especially at high metallicity, which indicates that self-shielding is important also in high-metallicity regions with $n_\mathrm{gas} < 10$ cm$^{-3}$. In the $f_\mathrm{CO}$--$Z$ relation, the tendency is similar to that at $t=$ 0.3 Gyr except that gas particles have higher metallicities. In the $X_\mathrm{CO}$--$Z$ relation, more gas particles have lower $X_\mathrm{CO}$ than at $t=$ 0.3 Gyr because of higher $f_\mathrm{CO}$. The majority of the gas particles still have $X_\mathrm{CO}$ larger than the theoretical prediction.

At the later stages such as $t=$ 5 and 10 Gyr, data points are much more concentrated at the higher end of metallicity. In the $f_\mathrm{H_2}$--$Z$ relation, many gas particles with $n_\mathrm{gas}$ < 10 cm$^{-3}$ have $f_\mathrm{H_2} > 0.1$. As metal enrichment proceeds, CO formation becomes more efficient because of more dust shielding. The correlation between $f_\mathrm{CO}$ and metallicity remains. At $t=$ 5--10 Gyr, all of the dense gas particles ($n_\mathrm{gas} > 10$ cm$^{-3}$) have $f_\mathrm{H_2} > 0.1$ and we observe that $X_\mathrm{CO}$ overall becomes lower than at earlier stages. Around the solar metallicity, $X_\mathrm{CO}$ is consistent with the Galactic value for a significant part of the gas particles, while we still find many gas particles with $X_\mathrm{CO}$ significantly higher than $X_\mathrm{CO, Gal}$. As explained above, the hydrogen column density calculated by $N_\mathrm{H} = n_\mathrm{H} h$ is used in our subgrid model, but it is probable that dense clouds, in reality, have a higher column density, which would lead to a higher CO abundance. Moreover, we completely neglected CO formation in gas particles with $n_\mathrm{gas} < 10$ cm$^{-3}$. Therefore, we have to keep in mind that our model might underestimate the CO abundance.
As mentioned in Section \ref{H2ab}, we also executed a simulation with a higher spatial resolution ($\epsilon_{\rm grav}$ $\sim$ 10$\,$pc) up to $t=1$ Gyr, but, because of the increase of small CO-rich clumps, we found more data points with low CO surface densities. Thus, we should note that $X_\mathrm{CO}$ does not necessarily decrease in a higher-resolution run.
In any case, resolving dense molecular clouds in a galaxy-scale simulation is not possible under the currently available computational power.
Another possible reason for underestimating the CO abundance is the underestimate of dust-to-gas ratio in our adopted simulation as mentioned in Section \ref{dust}.

\section{Comparison with observations -- star formation law}
\label{obs}
Observations of nearby galaxies have shown that the SFR surface density has a linear relation with the molecular gas surface density. We examine if our model can reproduce this star formation law. Objects for which the star formation law can be derived are biased to nearby metal-rich objects because CO is difficult to detect in metal-poor objects \citep[e.g.][]{Bolatto:2013aa}. Thus, we use the snapshots at 5 and 10 Gyr for the comparison with observations.

We adopt the observational data in \citet{Leroy:2013aa}. They estimated the spatially resolved molecular gas surface density using $^{12}$CO(2-1) maps, and derived $\Sigma_\mathrm{SFR}$ from maps of various star formation tracers (H$\alpha$, UV and 24 $\mu$m surface brightness) for 30 nearby disc galaxies with a common sub-kpc spatial resolution. They considered both the Galactic CO-to-H$_2$ conversion factor and a dust-to-gas ratio dependent one to convert the CO intensity to the surface density of molecular gas; however the resulting star formation law is overall similar between these two conversion factors and among various star formation tracers. Other studies also found a similar star formation law in nearby galaxies \citep{Wong:2002aa, Bigiel:2008aa, Schruba:2011aa, Rahman:2012aa}.

\begin{figure*}
	\begin{center}
	\includegraphics[width=2\columnwidth]{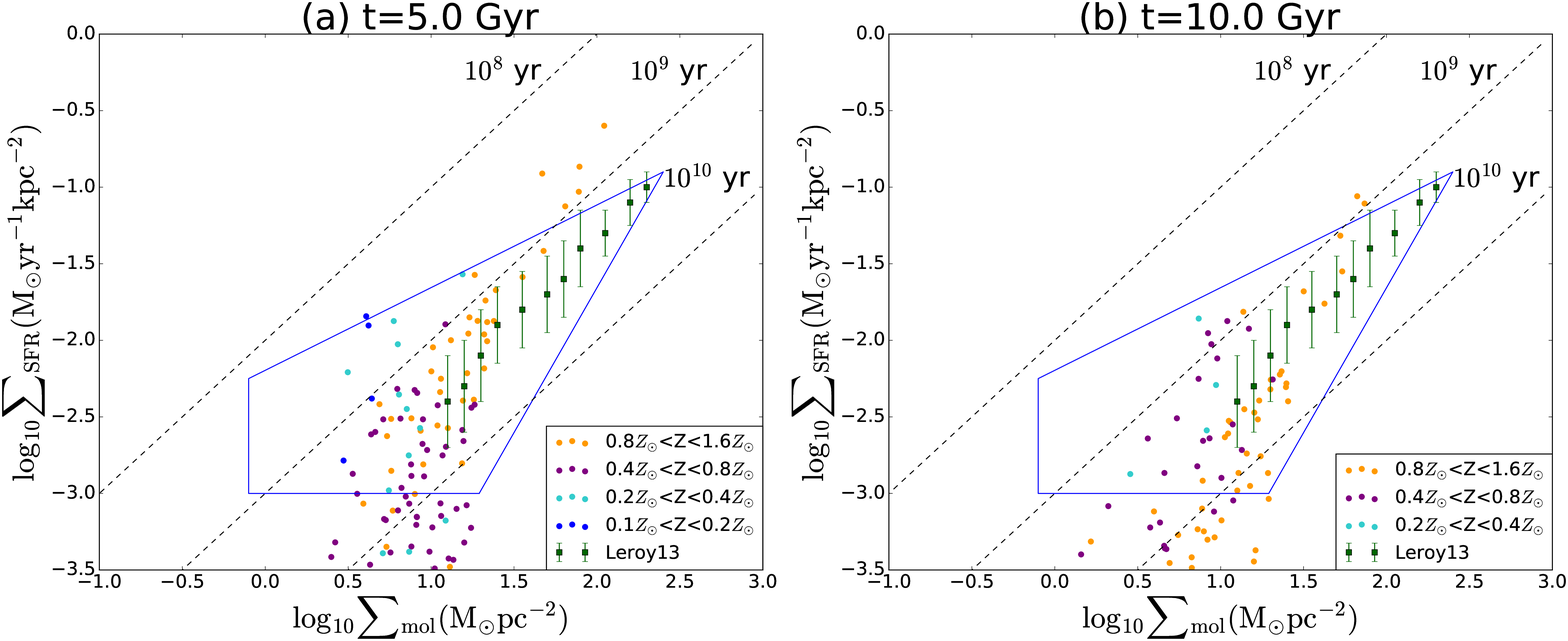}
	\caption{Star formation rate surface density, $\Sigma_\mathrm{SFR}$, from the simulation versus the molecular gas surface density $\Sigma_\mathrm{mol}$ calculated by our model. Each filled circle represents each grid point in the simulation map ($L_\mathrm{grid} = 0.5$kpc) and its colour indicates the metallicity as shown in the legend. The dotted lines show $\tau^\mathrm{mol}_\mathrm{dep} = 10^{8}$, $10^{9}$, and $10^{10}$ yr, from top to bottom. The area surrounded by the solid blue lines covers the distribution of observational data in \protect\citet{Leroy:2013aa} while the green squares and error bars show the running median and standard deviation.}
	\label{fig:obssflaw}
	\end{center}
\end{figure*}

\begin{figure*}
	\begin{center}
	\includegraphics[width=2\columnwidth]{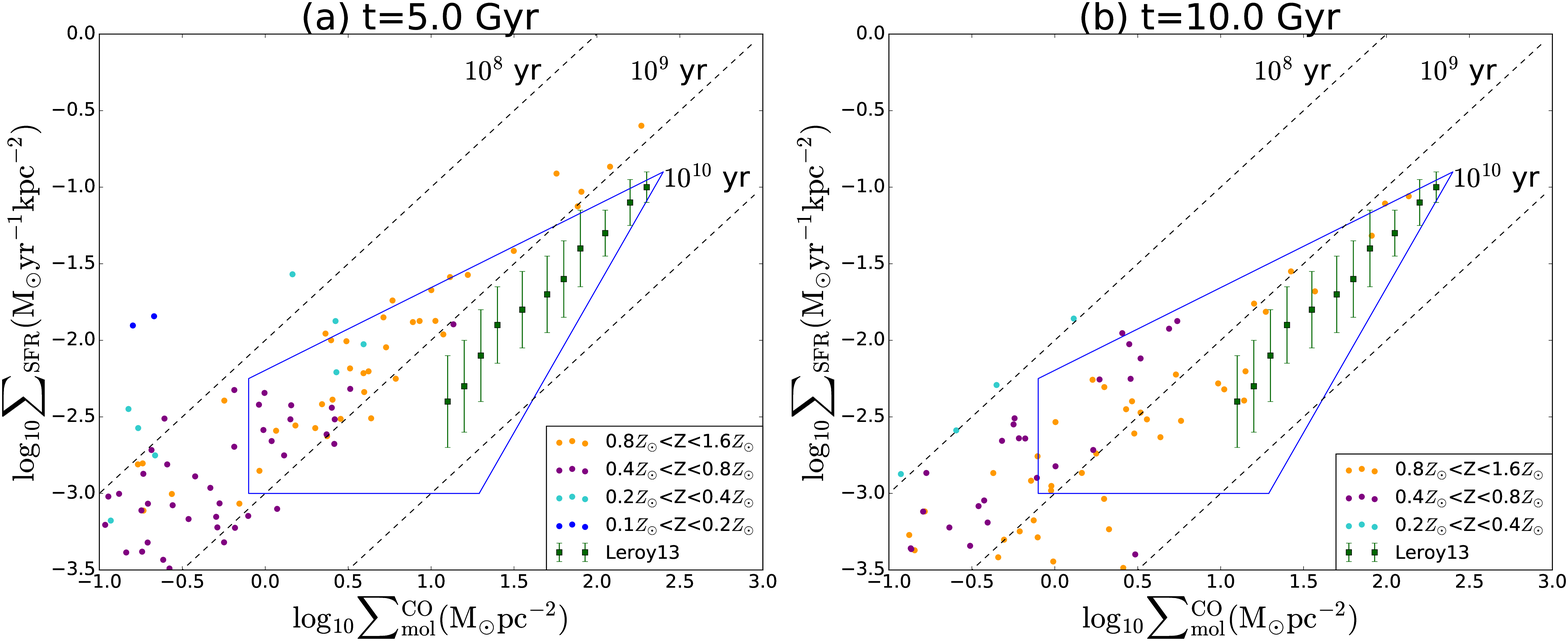}
	\caption{Same as Fig.~\ref{fig:obssflaw} but using $\Sigma^\mathrm{CO}_\mathrm{mol}$ (the molecular gas surface density is derived from CO using the Galactic conversion factor) in our model for abscissa. The dotted lines show $\tau^\mathrm{CO}_\mathrm{dep} = 10^{8}$, $10^{9}$, and $10^{10}$ yr, from top to bottom.}
	\label{fig:obssflawco}
	\end{center}
\end{figure*}

For our simulation data, we adopt $L_\mathrm{grid} = 0.5$ kpc, which is similar to the spatial resolution of the observational maps, for the smoothing (Section~\ref{grid}) and recalculate $\Sigma_\mathrm{mol}$, $\Sigma^\mathrm{CO}_\mathrm{mol}$, and $\Sigma_\mathrm{SFR}$. We show the resulting $\Sigma_\mathrm{SFR}$--$\Sigma_\mathrm{mol}$ and $\Sigma_\mathrm{SFR}$--$\Sigma^\mathrm{CO}_\mathrm{mol}$ relations in Figs.~\ref{fig:obssflaw} and~\ref{fig:obssflawco}, respectively. The observational data in \citet{Leroy:2013aa} is presented in the following two ways. The blue solid area covers the distribution of individual data points, while the green points show the running median with the standard deviations indicated by error bars. The observational data show a molecular gas depletion time of $\sim$ a few $\times~10^{9}$ yr. There is a cut-off of observational data of $\Sigma_\mathrm{SFR}$ at $\sim 10^{-3}$ M$_{\odot}$ yr$^{-1}$ kpc$^{-2}$ because of the observational detection limit. 

In Fig.~\ref{fig:obssflaw}, we present the $\Sigma_\mathrm{SFR}$--$\Sigma_\mathrm{mol}$ relation where the molecular gas abundance is estimated through H$_2$ abundance in our model (Section~\ref{H2ab}). The simulation data points are located along a constant depletion time broadly consistent with the observational data. However, at $t=$ 10 Gyr, there are only a few high--$\Sigma_\mathrm{mol}$ points. The lack of dense gas at $t \gtrsim$ 5 Gyr is already noted by A17, and is probably due to the consumption of gas by star formation. The $\Sigma_\mathrm{SFR}$--$\Sigma_\mathrm{mol}$ relations in our simulation shows a slightly steeper slope than the linear relation indicated by the running medians of observational data. 

In Fig.~\ref{fig:obssflawco}, we present the $\Sigma_\mathrm{SFR}$--$\Sigma^\mathrm{CO}_\mathrm{mol}$ relation. High-metallicity points at $t=$ 10 Gyr show roughly a consistent depletion time with observations. It is also remarkable that we clearly see an almost linear $\Sigma_\mathrm{SFR}$--$\Sigma^\mathrm{CO}_\mathrm{mol}$ relation at both ages with a flatter slope than in Fig~\ref{fig:obssflaw}. At $t=$ 10 Gyr, the points with solar metallicity in our simulation are consistent with the observational star formation law. However, a significant fraction of data points deviate towards lower $\Sigma^\mathrm{CO}_\mathrm{mol}$ compared with the observational data points especially at $t=$ 5 Gyr. This is probably because of the underestimation of CO abundance mentioned in Section~\ref{RZdep}. The difference between $t=$ 5 and 10 Gyr, can be explained by different ISRFs (i.e. different dissociation rates) with a stronger ISRF at $t=$ 5 Gyr than at $t=$ 10 Gyr because of higher star formation activity.


\section{Discussion}
\label{Discussion}
In Section~\ref{Results}, we showed the molecular gas and CO-based molecular gas abundances at different stages of galaxy evolution. At the early ($t \lesssim$ 1 Gyr) or low-metallicity ($Z~\lesssim~0.1~Z_{\odot}$) stage, the coupling between molecular gas surface density and star formation activity is weak because of inefficient molecular gas formation on dust; that is, molecular gas does not trace dense regions suitable for star formation. This means that, if we use molecular-based star formation model, the SFR is underestimated. In particular, CO formation is strongly suppressed compared with H$_2$ formation because it requires more shielded environment. At the later stage, as A17 mentioned, the dense gas is depleted. Nevertheless, CO formation is still enhanced due to the dust enrichment but the spatial distribution is limited because of the depletion of dense gas. At this stage, molecular gas abundance is tightly related to star formation activity because of efficient molecular gas formation.

However, the above behaviours of molecular abundances and star formation law may depend on some modelling features. In particular, since the spatial resolution in our simulation is limited, we applied the subgrid model to each gas particle as described in Section~\ref{H2ab}. Because the subgrid model is fundamentally important to the molecular gas abundance in our model, it is worth examining the effect of changing the subgrid parameters ($f_\mathrm{dense}$ and $n_\mathrm{H,d}$). The grain size distribution is also important for both H$_2$ formation and CO formation because it affects the dust optical depth (equation~\ref{eq:optical_depth}) and H$_2$ formation rate (equation~\ref{eq:reaction_rate}). Although the grain size distribution is consistently calculated with the hydrodynamic and chemical evolution of the ISM, it is still worth clarifying how much the grain size distribution affects the resulting H$_2$ and CO abundances. We discuss the dependence on the subgrid model and grain size distribution in the following subsections. 

\subsection{Variation of the subgrid parameters for H$_2$ formation}
\label{DH2ab}
\begin{figure*}
	\begin{center}
	\includegraphics[width=2\columnwidth]{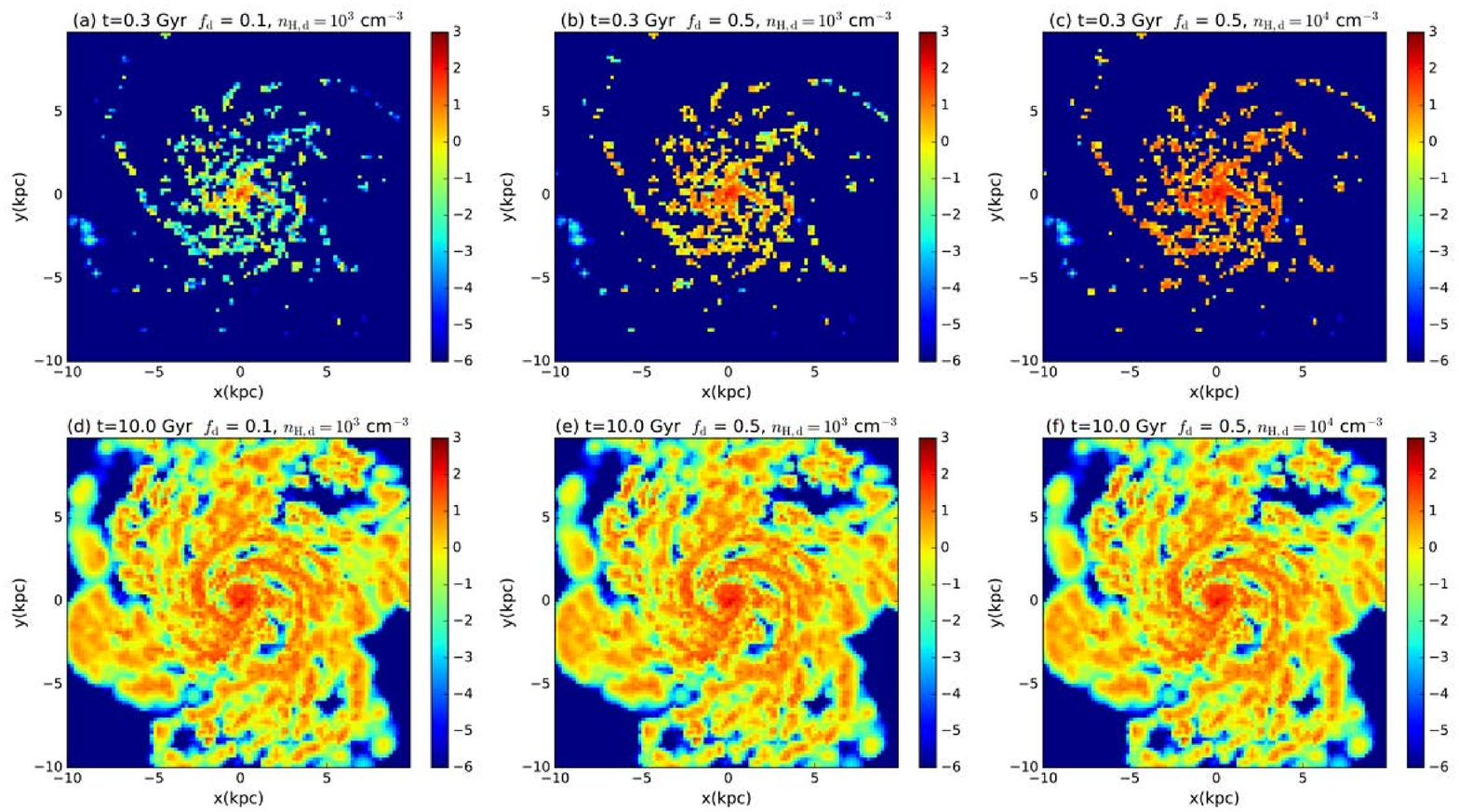}
	\caption{Surface density distribution of molecular gas with different parameter settings in the subgrid model. The upper and lower panels show the results at $t=$ 0.3 Gyr and $t=$ 10 Gyr, respectively. The hydrogen number density and the fraction of the subgrid dense clouds ($n_\mathrm{H,d}$, $f_\mathrm{dense}$) are chosen as (0.1, $10^3$ cm$^{-3}$), (0.5, $10^3$ cm$^{-3}$) (fiducial), and (0.5, $10^4$ cm$^{-3}$) for Panels a/d, b/e, and c/f, respectively. The logarithmic molecular gas surface density in units of M$_{\odot}$pc$^{-2}$ is indicated by the colour bar.}
	\label{fig:DsH2xy}
	\end{center}
\end{figure*}

We discuss the robustness of our subgrid model for H$_2$ formation formulated in Section~\ref{H2ab} (note that CO formation is not affected by the choice of $f_\mathrm{dense}$ and $n_\mathrm{H,d}$; see Section~\ref{COform}). We tuned two variables in the subgrid model: the mass fraction of dense clouds, $f_\mathrm{dense}$, and the number density of hydrogen inside the dense clouds, $n_\mathrm{H,d}$. Recall that we adopted $f_\mathrm{dense} = 0.5$ and $n_\mathrm{H,d} = 10^{3}$ cm$^{-3}$ for the fiducial values (Section~\ref{H2ab}). To examine how the predicted molecular gas surface density is affected by those parameters, we change either $f_\mathrm{dense}$ from 0.5 to 0.1 or $n_\mathrm{H,d}$ from $10^{3}$ to $10^{4}$ cm$^{-3}$. As representative ages for early metal-poor and late metal-rich phases, we show the results at $t=$ 0.3 and 10 Gyr, respectively in Fig.~\ref{fig:DsH2xy}.

At $t=$ 0.3 Gyr, we first examine the effect of tuning $f_\mathrm{dense}$ (Figs.~\ref{fig:DsH2xy}a and b). With $f_\mathrm{dense} = 0.5$, higher H$_2$ abundance is obtained as shown in Fig.~\ref{fig:DsH2xy}b. As expected, the molecular gas surface density is almost proportional to $f_\mathrm{dense}$. Next, we examine the effect of tuning $n_\mathrm{H,d}$ (Figs.~\ref{fig:DsH2xy}b and c). The H$_2$ formation is largely enhanced and the overall H$_2$ abundance is much higher with $n_\mathrm{H,d} = 10^{4}$ cm$^{-3}$. Thus, the absolute level of $\Sigma_\mathrm{mol}$ is strongly influenced by the subgrid model in the early, low-metallicity phase.

At $t=$ 10 Gyr, as we observe in Figs.~\ref{fig:DsH2xy}d -- f,  difference in the molecular gas surface density is not clear even between the two extreme choices of $f_\mathrm{dense}$ and $n_\mathrm{H,d}$ in contrast to the results at $t=$ 0.3 Gyr. We argue that these different behaviours at $t=$ 0.3 and 10 Gyr are caused by different dust abundance. At an early stage, when the galaxy is metal-poor, the H$_2$ fraction is $\ll$ 1 in most regions, having room for increase as well as decrease.  At a later stage, when the galaxy is metal-rich, H$_2$ formation is efficient enough to make the entire dense gas particles ($n_\mathrm{gas} > $ 10 cm$^{-3}$) fully molecular; thus, the subgrid treatment is not important for the molecular gas abundance. In other words, the current spatial resolution is enough to robustly predict $\Sigma_\mathrm{mol}$ in solar-metallicity environment.

\begin{figure*}
	\begin{center}
	\includegraphics[width=2\columnwidth]{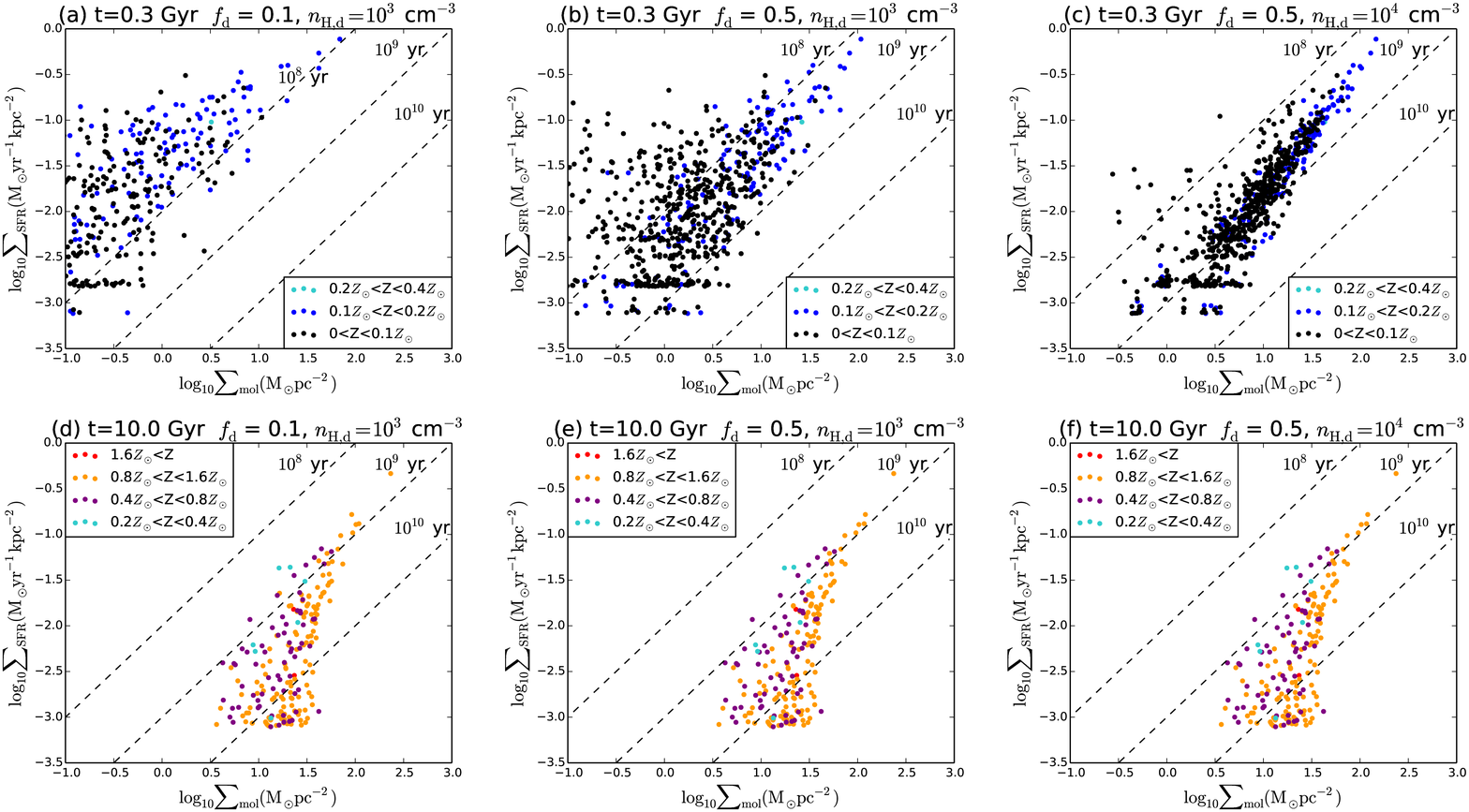}
	\caption{Star formation law with different parameter settings in the subgrid model. The upper and lower panels show the results at $t=$ 0.3 Gyr and $t=$ 10 Gyr respectively. The choice of the hydrogen number density and the fraction of subgrid dense clouds ($n_\mathrm{H,d}$ and $f_\mathrm{dense}$) is shown on each panel. The metallicity is colour-coded and lines with constant $\tau^\mathrm{mol}_\mathrm{dep}$ are shown in the same way as in Fig.~\ref{fig:RSFRH2}.}
	\label{fig:DsSFRH2}
	\end{center}
\end{figure*}

We also show the star formation laws for various $f_\mathrm{dense}$ and $n_\mathrm{H,d}$ in Fig.~\ref{fig:DsSFRH2}, where we use the molecular gas surface density directly (not through CO). At $t=$ 0.3 Gyr, if we increase $n_\mathrm{H,d}$, a tighter star formation law appears because low-metallicity points, which produce a scatter toward low molecular gas surface densities, shift toward high $\Sigma_\mathrm{mol}$ (tend to converge to a linear star formation law). The data points with high $\Sigma_\mathrm{mol}$ are less affected by the change of $n_\mathrm{H,d}$, since H$_2$ formation in the dense clouds is mostly saturated. When changing $f_\mathrm{dense}$ from 0.1 to 0.5, the data points overall shift toward higher molecular gas surface densities. This reflects the change in the fraction of molecular rich cloud (thus, the molecular gas abundance is almost proportional to $f_\mathrm{dense}$). At $t=$ 10 Gyr, no matter which parameter is changed, the distribution of data points on the star formation law diagram remains almost identical. This is because, as mentioned above, H$_2$ formation is efficient enough to realize $f_\mathrm{H_2} \sim 1$ in dense regions regardless of the detailed subgrid setting. We conclude that the small-scale (subgrid) structure is important for the molecular gas surface densities and the star formation law at low metallicity (or in the early phase of galaxy evolution) but is not important at high ($\sim$ solar) metallicity. At any age, the trend that lower-metallicity regions within the galaxy are deviated toward lower molecular gas surface densities on the star formation law diagram is commonly seen. 

\subsection{Effects of grain size distribution}
\label{Dsize}
In our dust evolution simulation, we calculate the evolution of grain size distribution in the form of large and small grain abundances. Thus, we refer to the small-to-large grain abundance ratio ($\mathcal{D}_\mathrm{S}$/$\mathcal{D}_\mathrm{L}$) as the grain size distribution in the following discussion. Here we discuss how the grain size distribution affects the H$_2$ and CO abundances. In order to clarify the importance of grain size distribution, we assume two extremes here: one is that all dust grains are large grains ($\mathcal{D}_\mathrm{L} = \mathcal{D}, \mathcal{D}_\mathrm{S} = 0$; referred to as the large grain case) and the other is that all dust grains are small grains ($\mathcal{D}_\mathrm{L} = 0, \mathcal{D}_\mathrm{S} = \mathcal{D}$; small grain case). Note that these are artificial assumptions for the post-processing calculations of the H$_2$ and CO abundances, and we do not change the dust model in the simulation. We compare these two extremes with the results in Sections~\ref{RH2ab}--\ref{RSFlaw}, where we used the grain size distribution given by the simulation (referred to as the standard case). Below we show the comparisons between the three cases of the grain size distribution.

\subsubsection{Molecular gas abundance}
\label{DH2}
\begin{figure*}
	\begin{center}
	\includegraphics[width=2\columnwidth]{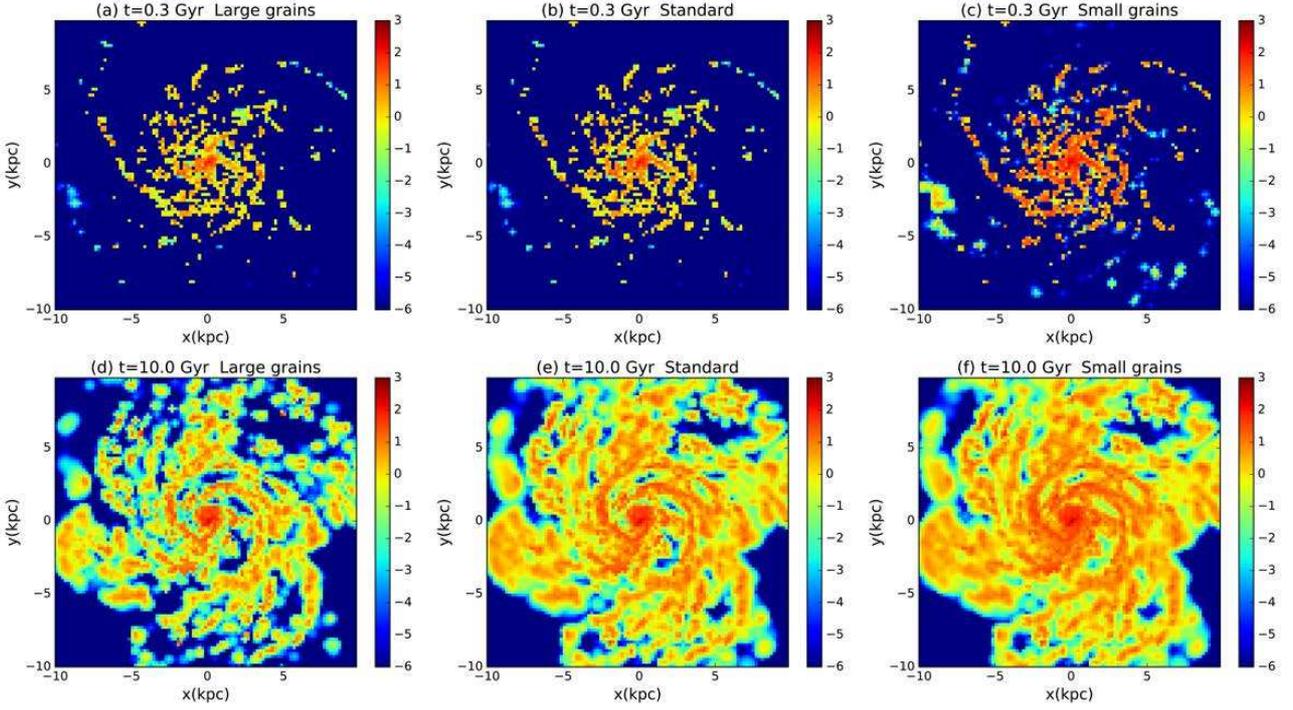}
	\caption{Surface density distribution of molecular gas ($\Sigma_\mathrm{mol}$) with different grain size distributions. The upper and lower panels show the results at $t=$ 0.3 and 10 Gyr, respectively. The left, middle, and right panels present the results for the large grain case, standard case, and small grain case, respectively. The colour levels (logarithmic surface density in units of M$_{\odot}$ pc$^{-2}$) are indicated by the colour bar.}
	\label{fig:DdH2xy}
	\end{center}
\end{figure*}

In Fig.~\ref{fig:DdH2xy}, we show the distribution of molecular gas surface density ($\Sigma_\mathrm{mol}$) for the small grain case, large grain case, and standard case at $t=$ 0.3 Gyr and $t=$ 10 Gyr. As formulated in Section~\ref{Model}, the formation rate of H$_2$ and the optical depth for dissociation radiation are inversely proportional to the dust grain size (equations~\ref{eq:reaction_rate} and \ref{eq:optical_depth}). Therefore, we expect that, if $\mathcal{D}_\mathrm{S} / \mathcal{D}_\mathrm{L}$ is large, the molecular abundances are enhanced.

We observe in Fig.~\ref{fig:DdH2xy} that at $t=$ 0.3 Gyr, the standard case (Fig.~\ref{fig:DdH2xy}b) is similar to the large grain case (Fig.~\ref{fig:DdH2xy}a), and that the small grain case  (Fig.~\ref{fig:DdH2xy}c) produces higher molecular gas abundance. This is because at the early stage, when most of the dust originates from stellar dust production, the dust is dominated by large grains. As the galaxy evolves, shattering and accretion increase the abundance of small grains in the standard case. Thus, at a later stage, dust in the galaxy is a mixture of large and small grains. At $t=$ 10 Gyr, the large grain case (Fig.~\ref{fig:DdH2xy}d) gives slightly less distributed molecular gas than the small grain case (Fig.~\ref{fig:DdH2xy}f) and the standard case shows the intermediate molecular gas surface densities (Fig.~\ref{fig:DdH2xy}e). However, the difference among the three cases in spiral arms is small because $f_\mathrm{H_2} \sim 1$ is realized in dense clouds (i.e. H$_2$ formation is saturated). The difference in $\Sigma_\mathrm{mol}$ among the three cases appears predominantly in the diffuse regions such as interarm regions. 

\begin{figure*}
	\begin{center}
	\includegraphics[width=2\columnwidth]{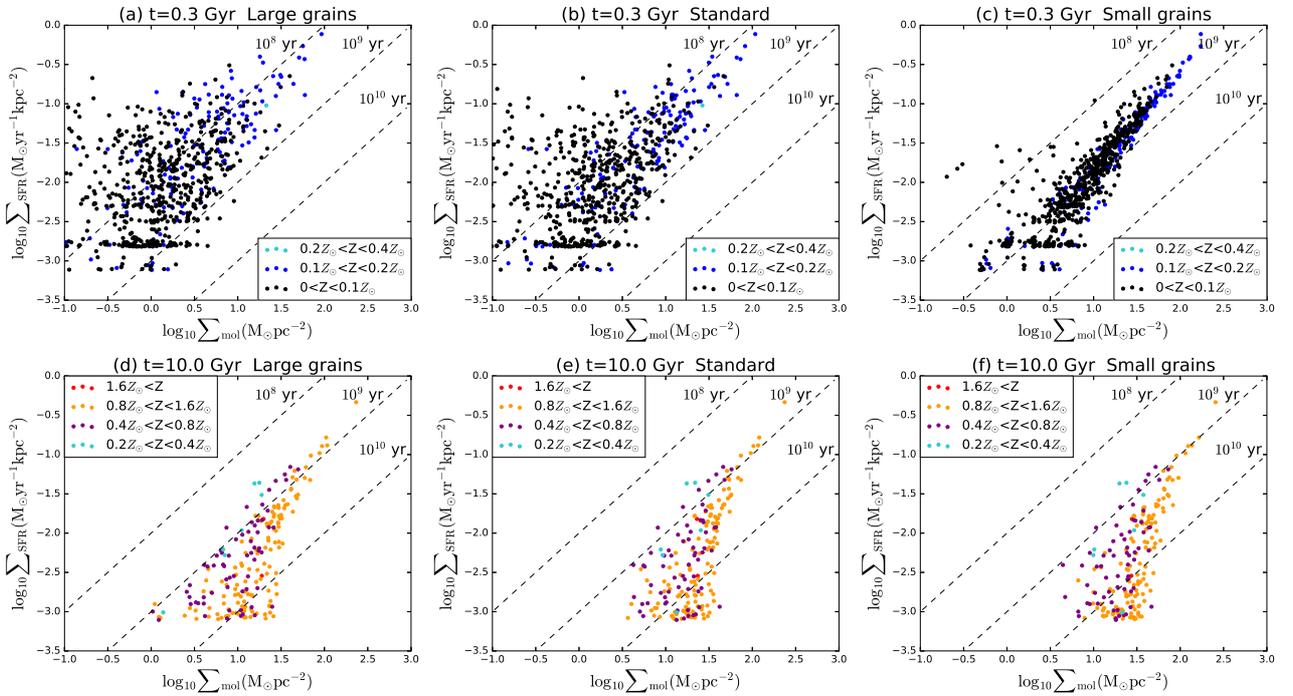}
	\caption{Star formation law ($\Sigma_\mathrm{SFR}$--$\Sigma_\mathrm{mol}$ relation) with different grain size distributions. The upper and lower panels show the results at $t=$ 0.3 and $t=$ 10 Gyr, respectively. The left, middle, and right panels present the results for the large grain case, standard case, and small grain case, respectively. The metallicity is colour-coded and constant $\tau^\mathrm{mol}_\mathrm{dep}$ is shown in the same way as in Fig.~\ref{fig:RSFRH2}.}
	\label{fig:DdSFRH2}
	\end{center}
\end{figure*}

We examine the star formation law for the three cases of grain size distribution at $t=$ 0.3 Gyr and $t=$ 10 Gyr in Fig.~\ref{fig:DdSFRH2}. At $t=$ 0.3 Gyr, the data points of our calculations are distributed similarly between the large grain case (Fig.~\ref{fig:DdSFRH2}a) and the standard case (Fig.~\ref{fig:DdSFRH2}b) while the small grain case (Fig.~\ref{fig:DdSFRH2}c) shows more concentrated distribution. Inefficient H$_2$ formation caused by large grain sizes produces the large scatter toward low $\Sigma_\mathrm{mol}$. In the small grain case, the scatter in the star formation law is largely underestimated. In contrast, the distributions of data points in all three cases at $t=$ 10 Gyr are similar (Figs.~\ref{fig:DdSFRH2}d--f). This is because the molecular gas abundances in dense star-forming regions are saturated with $f_\mathrm{H_2} \sim 1$ regardless of the grain size distribution. Therefore, the grain size distribution is important for the star formation law at low metallicity, while it is not important at high metallicity.

\subsubsection{CO abundance}
\label{DCO}
\begin{figure*}
	\begin{center}
	\includegraphics[width=2\columnwidth]{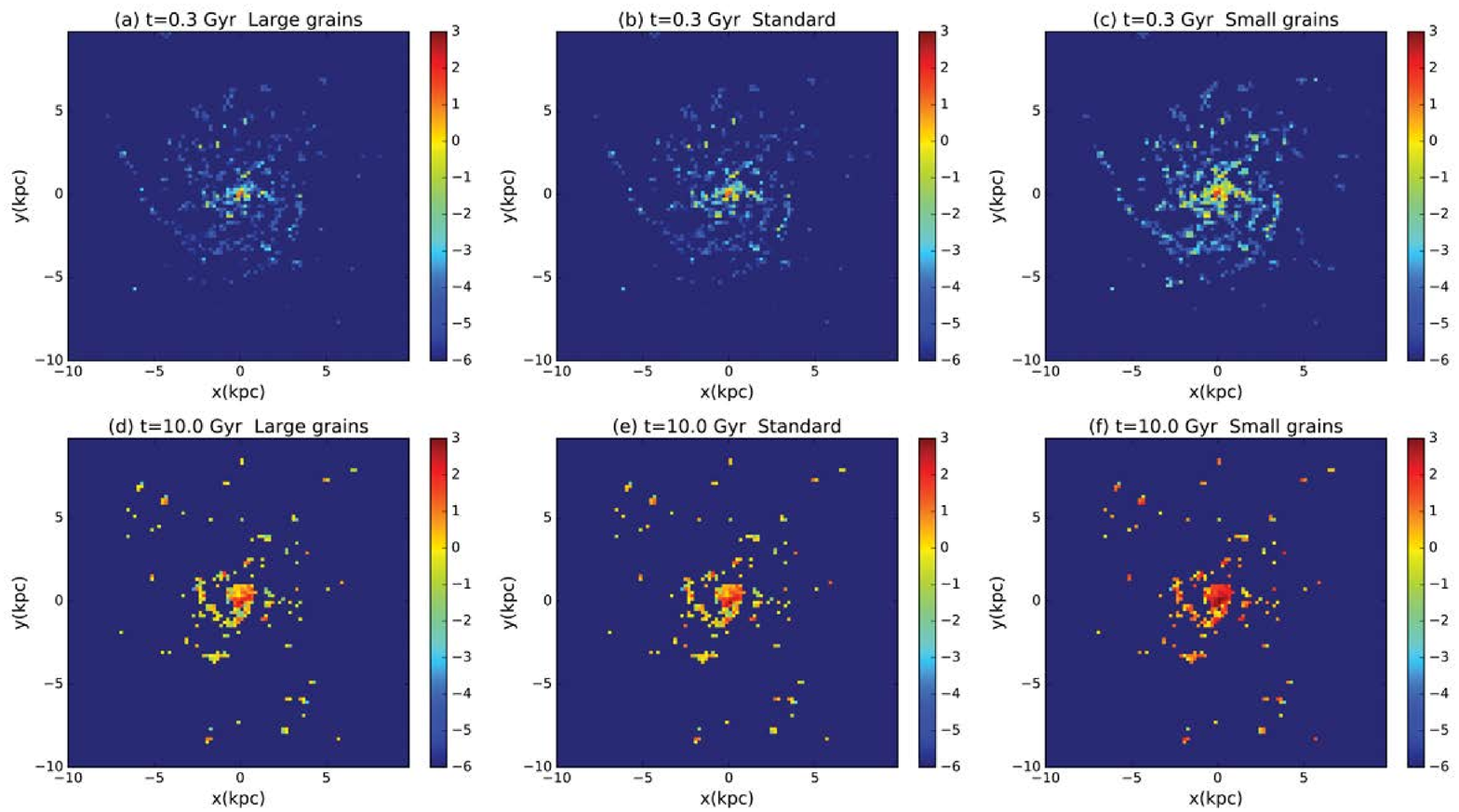}
	\caption{Same as Fig.~\ref{fig:DdH2xy} but using $\Sigma^\mathrm{CO}_\mathrm{mol}$ (the molecular gas surface density is derived from the CO abundance using the Galactic conversion factor).}
	\label{fig:Ddnew_H2xy}
	\end{center}
\end{figure*}

The grain size distribution affects not only H$_2$ formation but also CO formation through the shielding of dissociating radiation. In Fig.~\ref{fig:Ddnew_H2xy}, we show the spatial distribution of $\Sigma^\mathrm{CO}_\mathrm{mol}$ for the three cases of grain size distribution. At $t=$ 0.3 Gyr (Figs.~\ref{fig:Ddnew_H2xy}a--c), though CO formation is inefficient, the small grain case still gives higher CO abundance and wider spatial distribution than the other cases because of larger optical depth of dust (i.e. more efficient shielding of dissociating photons). As we mentioned in Section~\ref{DH2}, the results for the large grain case and the standard case are almost identical because the dust is dominated by large grains at this stage. At $t=$ 10 Gyr (Figs.~\ref{fig:Ddnew_H2xy}d--f), the CO abundance is higher in the small grain case than in the large grain case with the standard case predicting an intermediate abundance. Therefore, the grain size distribution is important for CO at both young and old ages (both low and high metallicities). The importance of small grains on the enhancement of CO abundance has already been suggested for damped Lyman $\alpha$ systems \citep{Shaw:2016aa, Fathivavsari:2017aa,Noterdaeme:2017aa}.

\begin{figure*}
	\begin{center}
	\includegraphics[width=2\columnwidth]{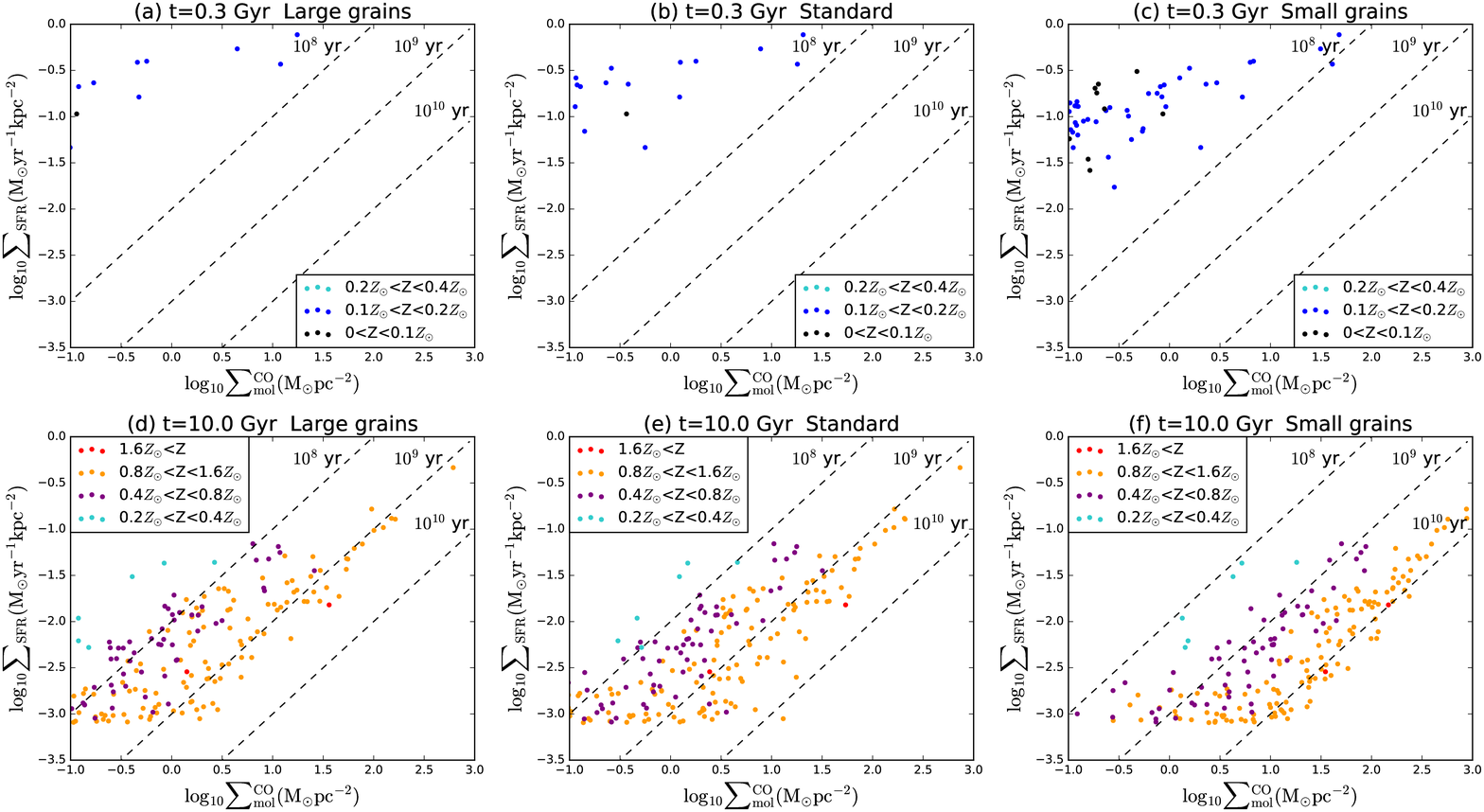}
	\caption{Same as Fig.~\ref{fig:DdSFRH2}, but using the CO-based values, $\Sigma^\mathrm{CO}_\mathrm{mol}$ and $\tau^\mathrm{CO}_\mathrm{dep}$.}
	\label{fig:DdSFRnewH2}
	\end{center}
\end{figure*}

In Fig.~\ref{fig:DdSFRnewH2}, we show the star formation law using $\Sigma^\mathrm{CO}_\mathrm{mol}$ (referred to as the CO-based star formation law). At both $t=$ 0.3 Gyr and $t=$ 10 Gyr, the small grain case (Figs.~\ref{fig:DdSFRnewH2}c and f) gives the highest CO abundance because of the largest shielding of dissociating radiation. Compared with the H$_2$ star formation law in Fig.~\ref{fig:DdSFRH2}, the CO-based star formation law is sensitive to the grain size distribution even at high metallicities or at old ages. While the H$_2$ abundance is saturated in metal-rich environment, CO is not; this is why $\Sigma^\mathrm{CO}_\mathrm{mol}$ is more sensitive to the grain size distribution than $\Sigma_\mathrm{mol}$.

In summary, from the above analysis of H$_2$ and CO, we find that the H$_2$ abundance is sensitive to the grain size distribution at low metallicity, while the CO abundance is sensitive to the grain size distribution both at low and high metallicities. The star formation law traced by H$_2$ is also affected by the grain size distribution at low metallicity, while the CO-based star formation law is still sensitive to the grain size distribution even at solar metallicity. Thus, the dependence of $\Sigma^\mathrm{CO}_\mathrm{mol}$ on the grain size distribution should be carefully considered if we estimate molecular gas mass from CO emission.

\subsection{Effect of dust evolution}
\label{Dde}

Other than the effect of grain size distribution, it would be useful to clarify the importance of calculating dust evolution.
In the simulation, dust evolution is strongly linked to metal enrichment; thus, in order to investigate the importance of dust evolution, we change the link between dust and metals in the model presented in this paper. Here, dust evolution includes the evolution of dust abundance and grain size distribution. 

In this subsection, we focus on the snapshot at $t=10$ Gyr. 
We test the following three cases here to clarify the importance of various aspects of dust evolution: (a) We use the dust-to-gas ratio, $\mathcal{D}$, calculated in the simulation but adopt the \citet[][hereafter, MRN]{Mathis:1977aa} grain size distribution \citep[$\mathcal{D}_\mathrm{S} = 0.3 \mathcal{D}$ and $\mathcal{D}_\mathrm{L} = 0.7 \mathcal{D}$;][]{Hirashita:2015aa} instead of $\mathcal{D}_\mathrm{S}/\mathcal{D}_\mathrm{L}$ calculated in the simulation. (b) We use dust-to-gas ratio linearly scaled with metallicity $\mathcal{D} = 0.2 Z$ with the MRN grain size distribution. The factor 0.2 is roughly the average of dust-to-metal ratios in the galactic disc at $t=10$ Gyr. (c) We use constant dust-to-gas ratio ($\mathcal{D} = 0.004$) with the MRN grain size distribution. The value of dust-to-gas ratio is roughly the average of $\mathcal{D}$ in the galactic disc at $t=10$ Gyr. We compare these three cases with (d) the standard case (using $\mathcal{D}_\mathrm{S}$ and $\mathcal{D}_\mathrm{L}$ calculated for each gas particle in the simulation).

We find that the H$_2$ surface density distribution in the galactic disc is almost identical among the above four cases.
The star formation laws ($\Sigma_\mathrm{SFR}$--$\Sigma_\mathrm{mol}$) are also almost identical among the four cases.
Therefore, the H$_2$ abundance is successfully modeled by a constant dust-to-gas ratio (or a constant dust-to-metal ratio) and the MRN grain size distribution (at $t = 10$ Gyr).
Note that this does not mean that such a model is applicable to other ages, since neither the MRN grain size distribution nor constant dust-to-gas (dust-to-metal) radio is applicable to the metal-poor phase.

\begin{figure*}
	\begin{center}
	\includegraphics[width=2\columnwidth]{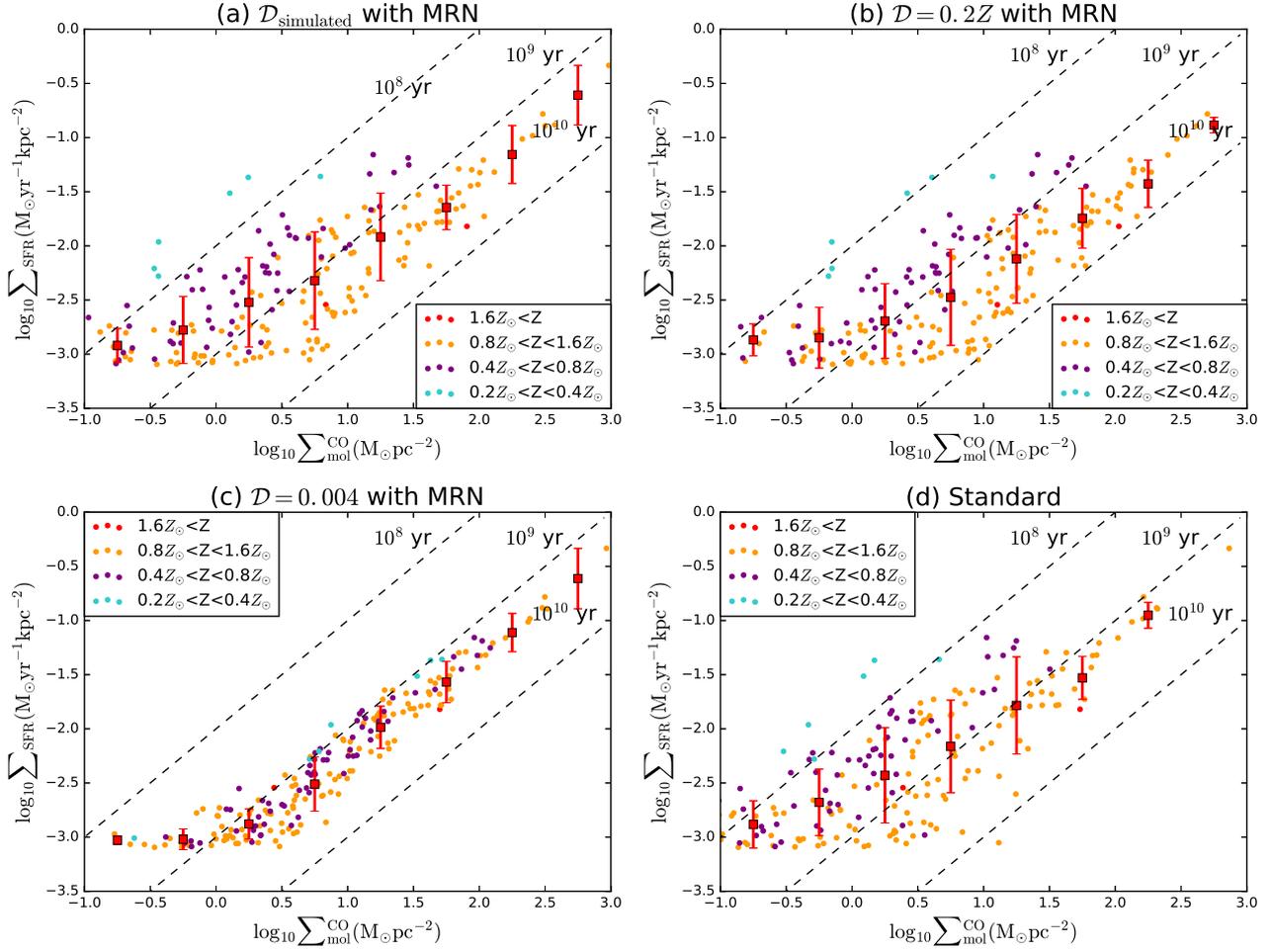}
	\caption{CO-based star formation law ($\Sigma_\mathrm{mol}^\mathrm{CO}$--$\Sigma_\mathrm{SFR}$ relation) at $t=10$ Gyr for the four different cases that are expected to give useful approximation to the metal-rich stage (like the Milky Way). Panel (a) shows the case in which we use the dust-to-gas ratio in the simulation but adopt the MRN grain size distribution. Panel (b) presents the case in which we adopt $\mathcal{D}=0.2Z$ with the MRN grain size distribution. In panel (c), we adopt $\mathcal{D}=0.004$ and the MRN grain size distribution, and in panel (d), we use $\mathcal{D}$ and $\mathcal{D}_\mathrm{S} / \mathcal{D}_\mathrm{L}$ calculated in the simulation. The colours of the points, the dashed lines, and the points with error bars are the same as in Fig.\ \ref{fig:RSFRH2}.
	}
	\label{fig:DSFRCO_DP}
	\end{center}
\end{figure*}

Because CO is confined in dense compact regions, it is not easy to see the difference in the spatial CO distributions in the above four cases. Thus, we examine the CO-based star formation law here.
In Fig.~\ref{fig:DSFRCO_DP}, we show the CO-based star formation law for the above four cases at $t=10$ Gyr. Case (a) with the MRN grain size distribution shows a similar CO-based star formation law to case (d). Thus, the MRN grain size distribution is good enough to predict the CO abundance in solar-metallicity environments. Case (b) ($\mathcal{D}=0.2Z$) also shows a similar relation to case (d); this indicates that adopting a reasonable value for the dust-to-metal ratio is essential to reproduce the CO abundance. Case (c) with a constant $\mathcal{D} = 0.004$, the data points show a significantly tighter relation than the other cases. Therefore, if we neglect the metallicity dependence of $\mathcal{D}$, we largely underestimate the dispersion of the CO-based star formation law.

In summary, we find that, at 10 Gyr (or in metal-rich environment),
CO formation is more sensitive to different dust properties than H$_2$ formation.
At such an evolved stage, adopting the MRN grain size distribution gives reasonable estimates of H$_2$ and CO abundances.
Adopting a constant dust-to-gas ratio gives a reasonable estimate for the H$_2$ abundance, but underestimates significantly the dispersion in the CO-based star formation law. Therefore, for CO, adopting an appropriate relation between dust-to-gas ratio and metallicity is fundamentally important.
We emphasize that it is essential to calculate the evolution of dust abundance and grain size distribution in understanding the star formation law at various evolutionary stages, since the above approximations are not valid at young (or low-metallicity) stages.

\section{Conclusions}
\label{Conclusions}
We investigate the spatial distribution of molecular gas (based on H$_2$ or CO) at different stages of an isolated disc galaxy using $N$-body/SPH simulation with dust evolution implemented. In the simulation, we calculate the evolution of dust abundance and grain size distribution (in the form of the large and small grain abundances) in a way consistent with the physical state of each gas particle in the simulation. H$_2$ formation on dust, UV dissociation, and self-shielding are included in the model by post-processing. We implement a subgrid model to take into account H$_2$ and CO formation in dense clouds ($n_\mathrm{H} \sim 10^{3}$ cm$^{-3}$), which are not resolvable by the simulation. We calculate the CO abundance in the subgrid model based on small-scale hydrodynamic simulation by GM11, who solved detailed chemical reactions. We assume that CO formation only occurs in dense clouds. The main results are as follows.
\begin{itemize}
\item H$_2$ formation is sensitive to the metallicity (or age). The H$_2$ surface density becomes higher and molecular-rich regions extend to the outer area of the galactic disc as the galaxy is enriched with dust. The star formation law is scattered at the early (metal-poor) stage and converges to a tight linear relation at the late (metal-rich) stage. The star formation law at $t\sim$ 5--10 Gyr predicts a molecular depletion time of a few Gyr, which is broadly consistent with the observational data. The large dispersion in the H$_2$ abundance at low metallicity implies that H$_2$ is a poor indicator of star formation site in the early phase of galaxy evolution; thus, the modelling of H$_2$-based star formation law needs careful implementation at low metallicity.
\item The CO surface density is more sensitive to the metallicity than the H$_2$ surface density; in particular, the CO abundance is strongly suppressed at low metallicity. We convert CO abundance to molecular abundance using the Galactic conversion factor and examine the CO-based star formation law. The CO-based star formation law also has a linear relation and is broadly consistent with the observation data at high metallicity; however, the star formation law in low-metallicity regions deviates from the observational star formation law.
\item We calculate the CO-to-H$_2$ conversion factor for each gas particle and confirm its negative dependence on metallicity, which is consistent with the observed trend. However, the individual CO-emitting regions are too compact to trace the overall H$_2$ distribution even at solar metallicity in our simulation, probably because our treatment only allows CO formation in dense gas particles, whose number declines rapidly at $t \sim$ 5--10 Gyr owing to gas consumption by star formation. Since our simulation does not resolve dense molecular clouds, we ignore CO formation inside those gas particles which are assumed not to have subresolution structure; the too compact CO distribution may be, to some extent, due to the limitation of our treatment, which could be resolved by future higher resolution simulations.
\end{itemize}

We further examine the dependence of H$_2$ abundance on the subgrid model in order to investigate the effect of structures below the spatial resolution of the simulation ($\sim$ a few tens pc). We find that the subgrid gas structures are more important at the early (metal-poor) stages, while at the later (metal-rich) stages, the H$_2$ surface density is insensitive to the subgrid model because hydrogen is mostly molecular in the relatively diffuse ($\sim$ 10--100 cm$^{-3}$) phase that can be resolved by the simulation. We confirm that the grain size distribution is important for CO formation since small grains shield UV radiation more efficiently than large grains. The H$_2$ abundance is sensitive to the grain size distribution in the low-metallicity phase while it is less sensitive to the grain size distribution at solar metallicity because the H$_2$ abundance is already saturated ($f_\mathrm{H_2} \sim 1$) in dense regions. At old ($\sim 10$ Gyr) ages, we find that adopting the MRN grain size distribution gives approximately correct H$_2$ and CO abundances. For the CO abundance, adopting an appropriate dust-to-metal ratio is important to correctly reproduce the dust evolution effect.

The current subgrid treatment assumes a fixed threshold to identify dense regions hosting dense clouds. Although the effects of metallicity evolution and grain size distribution are successfully investigated by our current framework, the subgrid model sets a limitation in predicting detailed CO and H$_2$ distributions. Further efforts are necessary to resolve this issue.

\section*{Acknowledgements}
We are grateful to the anonymous referee for useful comments. HH thanks the Ministry of Science and Technology for support through grant MOST 105-2112-M-001-027-MY3. KN, SA, and IS acknowledge the support from JSPS KAKENHI Grant Number JP26247022 and JP17H01111.




\bibliographystyle{mnras}
\bibliography{/Users/IcLa/Documents/ASIAA/reference.bib}







\bsp	
\label{lastpage}
\end{document}